\documentclass{sig-alternate}

\usepackage{multirow}
\usepackage{subfigure}
\usepackage{url}
\usepackage{comment}

\begin{document}
%
\conferenceinfo{CIKM}{'16, October 24-28, 2016, Indianapolis, USA}
\title{Searching for the Internet of Things on the Web: \\ Where It Is and What It Looks Like}

\numberofauthors{1}
\author{
\alignauthor Ali Shemshadi\textsuperscript{1}, Quan Z. Sheng\textsuperscript{1}, Wei Emma Zhang\textsuperscript{1},
Aixin Sun\textsuperscript{2}, \\Yongrui Qin\textsuperscript{3}, and Lina Yao\textsuperscript{4}
\\[2mm]
       \affaddr{\textsuperscript{1}School of Computer Science, The University of Adelaide, Australia
			}\\
       \affaddr{\textsuperscript{2}School of Computer Science and Engineering, Nanyang Technological University, Singapore
       }\\
       \affaddr{\textsuperscript{3}School of Computing and Engineering, University of Huddersfield, UK
       }\\
       \affaddr{\textsuperscript{4}School of Computer Science and Engineering, UNSW, Australia
       }\\       \vspace{0.02in}
       \email{\{ali.shemshadi,michael.sheng,wei.zhang01\}@adelaide.edu.au,\\
       axsun@ntu.edu.sg, y.qin2@hud.ac.uk, lina.yao@unsw.edu.au}
}

\maketitle
\begin{abstract}
The Internet of Things (IoT), in general, is a compelling paradigm that aims to connect everyday objects to the Internet. Nowadays, IoT is considered as one of the main technologies which contribute towards reshaping our daily lives in the next 
decade. 
IoT unlocks many exciting new opportunities in a variety of applications in research and industry domains. 
However, many have complained about the absence of the real-world IoT data. Unsurprisingly, a common question that arises regularly nowadays 
is ``{\em Does the IoT already exist?}". So far, little has been known about the real-world situation on IoT, its attributes, the presentation of data and user interests. To answer this question, in this work, we conduct an in-depth analytical investigation on real IoT data. 
More specifically, we identify IoT data sources over the Web and 
develop a crawler engine to collect large-scale real-world IoT data for the first time. We make the results of our work available to the public in order to assist the community in the future research. 
In particular, 
we collect the data of nearly two million Internet connected objects and study trends in IoT using a real-world query set from an IoT search engine.
Based on the collected data and our analysis, we identify the typical characteristics of IoT data. The most intriguing finding of our study is that IoT data is mainly disseminated using Web Mapping while the emerging IoT solutions such as the Web of Things, are currently not well adopted. On top of our findings, we further discuss future challenges and open research problems in the IoT area.
\end{abstract}
\vspace{-2mm}



\vspace{10mm}
\section{Introduction}
The Internet of Things (IoT) and IoT platform are two of the few concepts that are sitting at the top of the recent Gartner Hype Cycle\footnote{http://www.gartner.com/newsroom/id/3114217}.
Tremendous amount of considerations from many professionals 
place IoT as one of the top technologies that will revolutionize our daily lives over the next decade. 
Indeed, it is 
estimated that 
IoT would reach the so called Plateau of Productivity in Gartner Hype Cycle 
within 5-10 years.
The IoT paradigm can be applied in a variety of areas of applications including healthcare, mining industry, environmental sensing, transportation and logistics, and so on \cite{da2014internet,pang2013ecosystem}. For example, through the use of the IoT infrastructure, users can track the location and schedule of a certain aircraft in real-time\footnote{http://www.flightradar24.com}.
Various definitions of the IoT are traceable within the research community, and each of them targets at some strong interest to a specific type of applications or technologies. The very first definitions of IoT consider simple objects and RFID technology only. Later, IoT definitions broaden the purpose, perspective and the enabling technologies \cite{atzori2010internet}. In a broader sense, it is hard to limit the boundaries of IoT to specific applications or specific technologies. Thus, we envisage IoT as the set of initiations that publish the data generated by embedded and non-embedded sensors 
that are publicly available on the Web. 


The status of IoT is indeed similar to an iceberg. Due to its novelty, its visibility is still very limited to the extent that a common question for many people is that \textit{``Does the IoT already exist?"} \cite{want2015enabling}. Crawling and analyzing real-world IoT data may help to answer this question. In the context of World Wide Web, this is usually carried out by existing search engines such as Google. However, in the context of IoT, very little work has been carried out in this regard. To the best of our knowledge, the only working example of the IoT search engine is {\em Thingful}\footnote{http://www.thingful.net} and none of the IoT search engines in the literature have been deployed for real-world or large-scale data. Furthermore, the Thingful initiation itself is still limited and significant progress is needed to expand this area. 
One instance of such limitations is the public availability of the collected data. 
For example,
Thingful provides access to its 
data only via 
a dedicated UI. Another example of the limitations is the 
fast expiration of the data due to the highly dynamic nature of the IoT \cite{qinSFDWV2014,Qin-JNCA16}. \textit{Graph of Things}\footnote{http://graphofthings.org} is another interesting project which aims to provide live IoT data in real-time, which is still limited and can be potentially expanded in terms of scope and capabilities.

There is another search engine, namely Shodan\footnote{https://www.shodan.io} which also claims to be a search engine for IoT. The main difference between Shodan and IoT search engines such as Thingful, is that Shodan is basically designed as a search engine for hackers. It identifies and hacks into password protected devices connected to the Internet. Servers and routers as well as other Internet-connected devices have been archived with their IP addresses in 
its database. The website itself does not process sensor outputs. Due to its large and broad scope, catching everyday objects on this website is still difficult while servers and network devices constitute the majority of the things in its database. 
Due to ethical issues and scope matters, we do not include Shodan in our study.

Given the lack of powerful IoT search engines and the unavailability of large-scale IoT data, the visibility of IoT and its data is far from satisfying. This creates notable gaps in the IoT research and development \cite{want2015enabling} and 
still leaves many questions without answers. We list some of them as follows:
\begin{enumerate}
\item Does IoT already exist on the Web?
\item What technologies are used to make IoT visible?
\item Which aspects of IoT are more interesting to users? 
\item What are the characteristics of the large-scale IoT data?
\end{enumerate}

In this paper, we conduct an extensive study on the current status of the real-world IoT. Our main contributions are summarized as follows:

\begin{itemize}
\item We identify and classify IoT data sources into three categorizations, including {\em Cloud based IoT platforms}, {\em Web of Things enabled platforms}, and {\em Web Mapping}. 
Our practical experience 
provides strong evidence that IoT does exist on the Web nowadays and we suggest that more IoT research efforts are needed to take advantage of this availability. 

\item We design and implement a novel IoT crawling platform, named ThingSeek, to collect and analyze IoT data. ThingSeek is composed of a crawler and a visualization engine. Using the crawler,
we capture publicly available data from the major IoT data sources that we identify over the Web. We make the collected IoT dataset available to the 
public in order to boost the research related to IoT. 

\item We study the general user interests on IoT data by using a real world query log dataset from an IoT search engine. 
%
We also analyze the characteristics of the collected IoT data including spatio-temporal distributions of things, data dynamics, and data quality. 

\item Based on the collected real-world IoT data and our analysis, we discuss future research challenges and identify open research problems to shed light on the future IoT research and development.
\end{itemize}

The rest of this paper is organized as follows.
We discuss the potential places to look for IoT over the Web in Section \ref{sec:methodology}. 
In Section \ref{sec:crawler}, we discuss the best practices that we learn in IoT data acquisition. Then in Section \ref{sec:results}, we present the analytical results of the collected IoT data. 
We discuss some of the opportunities for further IoT research in Section \ref{sec:discussion}. In Section \ref{sec:relatedwork}, we overview the related works and Section \ref{sec:conclusion} concludes the paper.

\section{Where is the IoT?}
\label{sec:methodology}
The interactions 
with IoT 
can be realized in Machine-to-Machine (M2M) as well as Machine-to-Human (M2H) \cite{wu2011m2m}. The M2M approach is mainly used for smart things and enabled by predefined APIs, e.g., RESTful APIs \cite{kim2014m2m,castro2012analysis}. In contrast, M2H can include almost every object that 
are connected to the Internet and 
enabled using current Web protocols and existing IoT middleware.  Pioneering IoT cloud services such as Xively\footnote{www.xively.com}, Paraimpu~\cite{pintus2012paraimpu}, ThingSpeak\footnote{www.thingspeak.com/} and Sen.se\footnote{https://sen.se/mother} are some of the examples of IoT dedicated cloud services which provide infrastructure to store and share things data for various types of sensors. Nowadays, there are numerous examples of websites which focus on a specific type of applications such as tracking aircrafts\footnote{flightradar24.com}, marine traffic\footnote{www.marinetraffic.com}, traffic jams\footnote{www.waze.com/livemap} or Raspberry Pi board\footnote{rastrack.co.uk}. In the rest of this 
paper, we refer to these two types of 
IoT services as general and niche IoT services, respectively.  

In our work, we categorize IoT data sources into three groups, namely the {\em cloud based IoT platforms}, the {\em WoT enabled platforms}, and the {\em Web Mapping enabled data sources}. 

\subsection{Cloud Based IoT Platforms}
Cloud computing is a very popular demand-based Internet computing paradigm in which, shared resources, data and information are provided to computers and other devices on-demand. Since the introduction of the concept, numerous cloud services have been launched where each service is designed specifically for certain applications such as web hosting, file sharing, programming and etc. 

With the growth of the idea of connecting things to the Internet, one of the dominant visions for developing the IoT is to use cloud computing technologies to develop cloud services which facilitate the utilities for storing, sharing and visualizing IoT data through the conventional tools of the World Wide Web \cite{gubbi2013internet, jiang2014iot}. For this purpose, 
many services 
have been developed where Table \ref{tab:cloudservices} enlists some of them.

One of the main features of this category, is that the platforms have been designed with the idea of enabling any object of any kind to be connected to the IoT rather than being a solution designed specifically for a certain application. Furthermore, the service model of the cloud platforms, can provide more details about cloud based platforms for IoT.
Infrastructure as a Service (Iaas), Platform as a Service (Paas) and Software as a Service (SaaS) are the prevalent service models for cloud services. These models provide services at different levels from basic access to infrastructure to complete service via online application software and database, respectively. IoT cloud platforms in this category, such as the services in Table \ref{tab:cloudservices}, may follow a SaaS model for connecting devices to IoT. All of the mentioned services provide dashboard, API, 
M2M communication, middleware and the infrastructure to facilitate the IoT connection of the devices. In addition, many of the popular cloud services which follow other models, such as Amazon Web Services and Google Cloud, have recently provided tools for IoT integration. 

Basically, cloud platforms for heterogeneous IoT devices are considered as the primary means of sharing IoT data \cite{gubbi2013internet}. The data ownership policy of these platforms 
is 
 typically
set to be private with a few exceptions such as Xively that provides a public sharing option.
Public IoT data, if available, can be generally retrieved through a pre-designed API such as Xively API\footnote{https://personal.xively.com/dev/docs/api/}.

\begin{table}[]
\centering
\caption{Examples of IoT cloud services}
\label{tab:cloudservices}
\begin{small}
\begin{tabular}{|l|p{.65\linewidth}|}
\hline
\multicolumn{1}{|c|}{\textbf{Platform}} & \multicolumn{1}{c|}{\textbf{Description}}                \\ \hline
Xively                                       & Popular IoT cloud with open data                   \\ \hline
Paraimpu                                     & Social network with IoT cloud                            \\ \hline
TheThings.io                                 & IoT cloud with open data       \\ \hline
Ayla Networks                                & IoT cloud with smartphone app                    \\ \hline
Jasper                                       & Scalable IoT cloud with predefined business applications \\ \hline
Cloud Plugs                                  & IoT cloud with variety of development libraries          \\ \hline
ThingSpeak                                   & One of the earliest IoT clouds                   \\ \hline
Covisint                                     & Enterprise purpose built IoT cloud                       \\ \hline
particle.io                                  & IoT cloud with hardware development kits                 \\ \hline
ThingWorx                                    & IoT cloud with machine learning                 \\ \hline
\end{tabular}
\end{small}
\end{table}

\subsection{WoT Enabled Platforms}
The WoT concept describes approaches, frameworks and programming patterns that allow things to share their data through the World Wide Web. Currently, WoT is an active research area with a range of challenges and opportunities including security, resilience, intent oriented search, legal implications and so on \cite{raggett2015web}. Backed by existing WoT packages\footnote{https://github.com/webofthings}, these data sources create mashups to publish IoT data. One of the most popular WoT packages is the WoTKit \cite{blackstock2012iot}. Although some WoT packages have been used by IoT cloud services, we distinguish them from other cloud services (e.g., Xively) that are not developed based on the WoT. 
WoT can be applied in both of the traditional server (such as WeIO examples\footnote{//github.com/nodesign/weio/tree/master/examples}) and the cloud based (such as SenseTecnic\footnote{https://wotkit.sensetecnic.com/wotkit/}) environments. 

\subsection{Web Mapping Enabled Data Sources}
Web Mapping is the process of using online 
maps to browse and visualize geospatial data in a Web environment (e.g., Google Maps) \cite{haklay2008web}.
Web Mapping is more than just Web cartography.
There exist a wide variety of use cases for Web Mapping presentation of the data. 
In fact, we realize that a considerable number of Web pages with maps are providing IoT data and thus, include them in our list of data sources. The main categories of such data sources are as follows:
\begin{itemize}
\item {\em Real-time Transportation Information Services}: Real-time tracking services (e.g., FlightRadar24\footnote{http://flightradar24.com} and Arrivebus\footnote{http://www.arrivabus.co.uk/journeyplanner/help/en?\\tpl=livemap}) are designed to process and share the coordination of public transport services generated by embedded GPS devices. Unlike IoT cloud platforms, these services are often publicly available and data is visualized via Web Mapping. The most dynamic attributes of the objects in these networks are location-related including latitude and longitude. 

\item {\em Urban Crowdsensing Services}: Urban crowdsourcing services provide a platform for people to report and share their observations of things around them. For example, Waze\footnote{https://www.waze.com} provides a mobile phone application for users to report their locations, traffic jams, roadworks or police attendances. Although the collected data from this type of platforms is not originated from embedded physical sensors, the information is still related to physical or virtual things that people observe around themselves. Most often, the data is available through a Web based map. 

\item {\em Public Environmental Sensing Services}: These services include platforms that share the data originated by environmental sensors such as weather stations and pollution metrics. The data is available through a Web based map interface available to public. 
\end{itemize}

\vspace{1mm}
\section{IoT Data Acquisition}
\label{sec:crawler}
In this section we provide details on identifying the data sources of things  and collecting the things dataset.
We focus on the general idea of the IoT which is composed of two main words: \textit{Internet} and \textit{Things}. As every object occupies a space and the location is one of the key features of things, we limit the scope of our search to the sources which necessarily contain, or at least consider, the location data. The location data is usually represented as a unique tuple $(latitude,longitude)$ such that $latitude \in [-90,90]$ and $longitude \in [-180,180]$. 
On top of the location information, we also need to focus on specific features to identify the relevant sources of things data. 
We include some of the available WoT and IoT platforms to our search queries to cover popular IoT implementation techniques. We also consider Web Mapping as an IoT data sharing interface to identify more IoT data sources. 
Thus, we limit our search scope to the sources which contain a map. 

\begin{figure*}[!tb]
\centering
\subfigure []
    {\includegraphics[width=.65\linewidth]
    {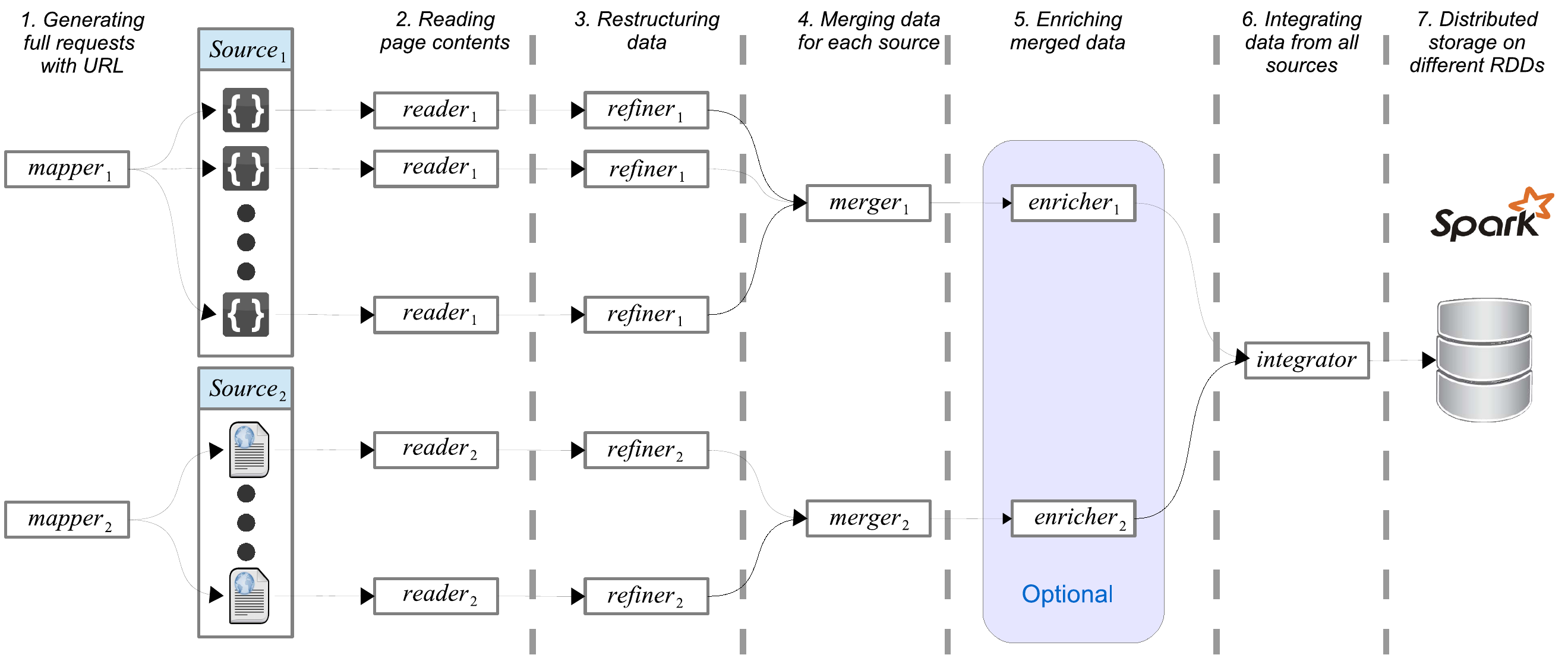}\label{fig:crawler}}
    \hspace{10mm}
\subfigure []
    {\includegraphics[width=.28\linewidth]
    {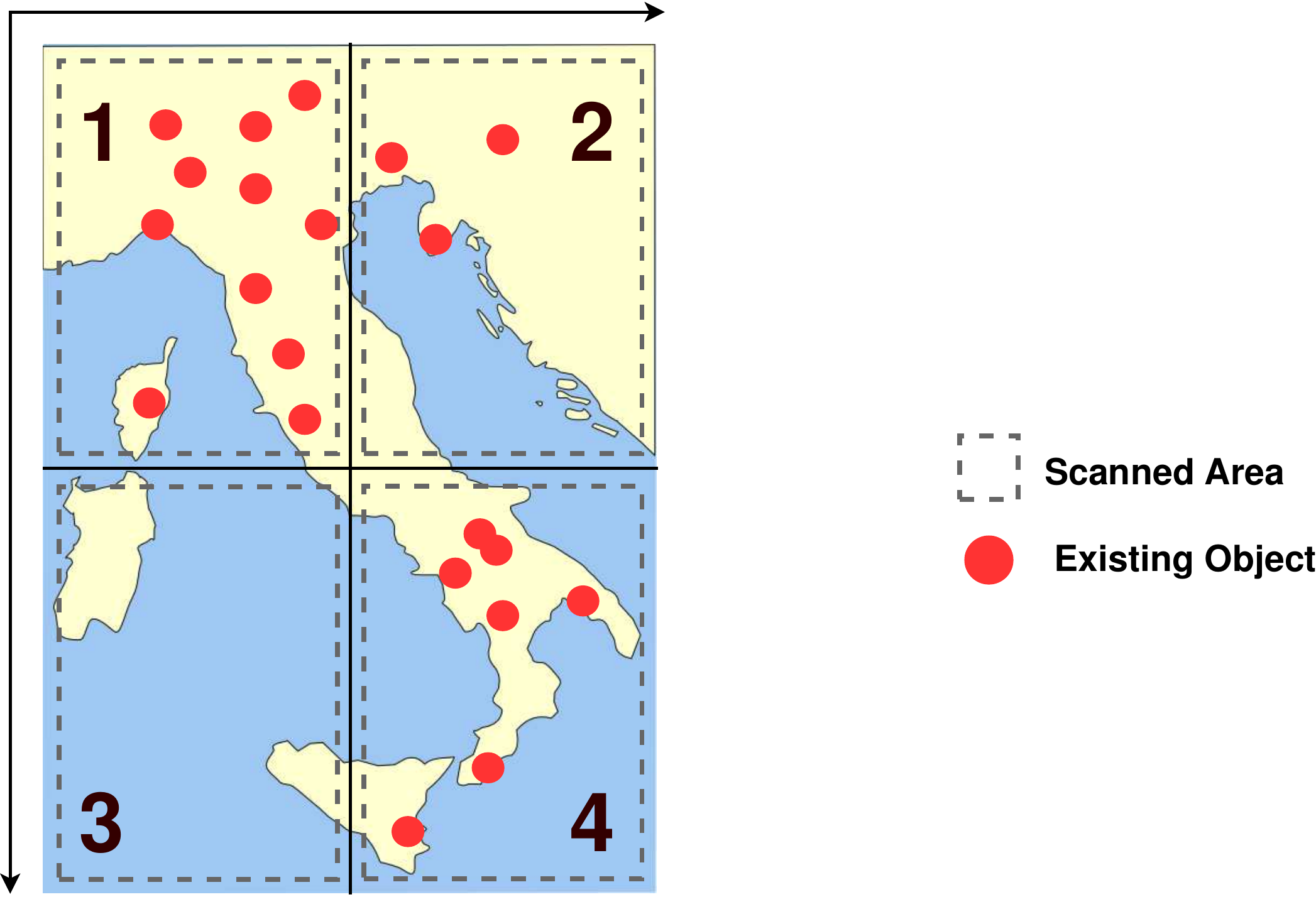}\label{fig:sequentialscan}}
    \caption{Illustration of (a) the ThingSeek crawler engine; and (b) sequential-spatial access to things data}
    \vspace{-2mm}
\end{figure*}

We break down the crawling procedure into a certain set of steps in a unified framework.
Figure \ref{fig:crawler} illustrates the main features of the ThingSeek crawler engine. In the first step, a URL generator initializes the queue of queries. Each entry in the queue is supplied with certain parameters to construct a query to a page or a specific location. The parameters can be the time window, the boundaries of the querying region and/or other parameters. Then, for each entity in the queue, a reader scans the page, and the contents are converted to a set of vectors and filtered by a refiner. The data for each subset is separately held until all subsets are refined where we merge all of the subsets of the resource's data. In this step, a specific enricher can be used to collect the missing information, if any, from other sources. This 
includes, for example, filling the incomplete fields such as IP address by acquiring them from Shodan. Finally, the collected data from different sources are integrated and stored on a distributed backend. 


Due to the size and dynamics of the sensor-generated data, IoT data sources often provide a subset of their data with a call to their API. Thus, pagination techniques such as location-based queries are deployed to present the data. We use the same mechanism through implementing the URL generator.
The URL generator plays a key role in adjusting the workload on the data source. It converts a set of spatial segments to a sequence of queries which can be submitted via the API of the data source. Thus, a highly populated area can be placed multiple times in the processing queue while an empty area may appear only once (or not appear) in the queue. 
We develop ThingSeek using a set of tools to collect, process and visualize the dataset. We use open source tools and libraries to implement our framework. Tools 
include R programming language, SparkR, Apache Spark 1.4.1 and Rails framework. 
\subsection{Identification of Data Sources}
The number of cloud IoT platforms with open access data is limited and thus, identifying them is not difficult. For WoT enabled data sources, one can check the traces of existing WoT packages such as the ones from 
WeIO\footnote{http://we-io.net}, WoT Code Forge\footnote{https://github.com/webofthings} and WoT Project Directory\footnote{https://github.com/webofthings/webofthings-projects-directory/wiki/Web-of-Things-Projects-Directory}.

In principle, not all IoT data appears in the form of Web Mapping and not every Web based map is related to IoT data. 
Web based maps have been used for a variety of purposes including presenting IoT data. 
From our experience, those Web pages that visualize IoT data have the following requirements: 1) containing an interactive map; 2) being publicly available; 3) being real-time; 4) being real-world; and 5) being within valid ranges.


IoT is usually updated in real-time and vintage maps are not very useful in this case. The real-world data is a key to find real physical things, thus, maps of virtual worlds such as game maps do not provide IoT data. Finally, key features of the data should contain proper values. Maps with encrypted data cannot be very useful for IoT data collection.

Based on the features of IoT data, which are mentioned above, we use the following procedure to identify the data sources: 1) the Web page should contain an interactive map; 2) data is presented inside the XMLHttpRequest (XHR) response of the requests that the page makes; 3) the response in the XHR may continuously be updated; and 4) data contains coordinates which are within the valid boundaries.

\subsection{IoT Data Collection}

We design a distributed crawler which can automate the process of IoT data collection. Firstly, we identify a set of potential sources and categorize them based on the pre-specified features and criteria. Then from the result set, we select 20 data sources from which many are partially included in Thingsful as well. As we need to further process Thingful queries with our crawled things dataset, the sources are selected to represent the Thingful data.
We crawl the selected data sources in two hour intervals for a period of one week between 25/8/2015 and 1/9/2015. The crawling resulted in two million things from the selected IoT data sources. 
Shortly, we release the collected things dataset for which a subset
is available online\footnote{http://tinyurl.com/z4d6aa6}. 
We distribute the crawler over 4 machines where each machine has the maximum of 2.5 GHz Intel core-i5 CPU, 8 GB memory. 

We observe that the majority of data sources use pagination to limit the output size. Thus, capturing the whole dataset through a single request 
using API would be impractical. For example, the length of the resultset of a query from a Web Mapping enabled website such as a flight tracker would be limited to a few hundred aircrafts. For this reason, we need to construct a spacial or paginated query and process it through the API. As shown in Figure \ref{fig:sequentialscan}, a larger area will be segmented into a grid of smaller segments and then we capture things data in a sub-area of each segment. In this regards, small margin will be considered to avoid capturing duplicate things. Duplicate things will be created if a single object is captured twice as it is moving from one segment to its neighbour.
However, using the above technique will result in sequential access to the whole data. In the example shown in Figure \ref{fig:sequentialscan}, Segment 4 may only be accessed after screening the other segments such as 1,2 and 3. Considering the high rate of updates as well as the size of data, this could make a problem as other segments such as 2 and 3 are not as populated as Segment 4. Thus, a considerable amount of resources is not used properly and a long delay in data refreshing is triggered by this mechanism.

Due to the frequent sensor reading updates, the volume of IoT data can be huge. Based on our observation, storing IoT data requires more than 0.25 GB of space per second. Thus, we can estimate that to store the data in 24 hours, we need approximately 21 TB of space. Obtaining this amount of information from a data source can be time consuming and bear a high cost. However, in order to save the processing cost and reduce the distortion of the captured things dataset, one option is to analyze the density of things in each area and put more effort on the areas with more things. For instance, in Figure \ref{fig:sequentialscan}, Segment 3 on average has the least density of things, thus it can be removed from the queue. Another option, which can be taken at the same time, is to consider the distribution of the retrieval queries. For instance in Figure \ref{fig:sequentialscan}, if user queries specify the Segment 2 more than Segment 4, more resources can be devoted to scan Segment 2 data.

In our dataset, we have identified nearly two million objects. The number of records for each object deviates between 10 and 1,933 based on the fact that the time for scanning different data sources varies between 30 seconds to more than 50 minutes. 
In 
average,
 the readings per object in our data set are 32.
%
%
%
On top of the crawled things dataset, we use a real-world query set consisting of 136,746 queries between 2/12/2014 to 1/27/2015 from the Thingful search engine. We use the query set to investigate user interests.
A query in our query set is structured as follows:
\begin{small}
\texttt{
\{"timestamp":"2015-01-27T09:33:06+00:00",  "query":\{"lat":\\"51.55", "lng":".03", "zoom":"8", "what":"speed camera"\}\}
}
\end{small}

%

%

\section{IoT Data Analysis}
\label{sec:results}
In this section, we present the result and statistical analysis of IoT data and queries collected during multiple routine crawling rounds. We investigate the results from user-related and things-related points of view and then we compare the distribution of IoT and queries data.

\subsection{User Interests}
We investigate user interests from different angles including \textit{Popularity Trends}, \textit{Search Queries Statistics} and comparing the \textit{Things vs Query Distribution}.

\subsubsection{Popularity Trends}
A glimpse into IoT keyword trends over Google Trends\footnote{http://trends.google.com/}, 
suggests that the public interest towards IoT with its most popular abbreviations has been steadily increasing over the past few years.

\begin{figure}[!tb]
\vspace{-1mm}
\centering
\includegraphics[width=.8\linewidth,trim={0 10 0 1}, clip]{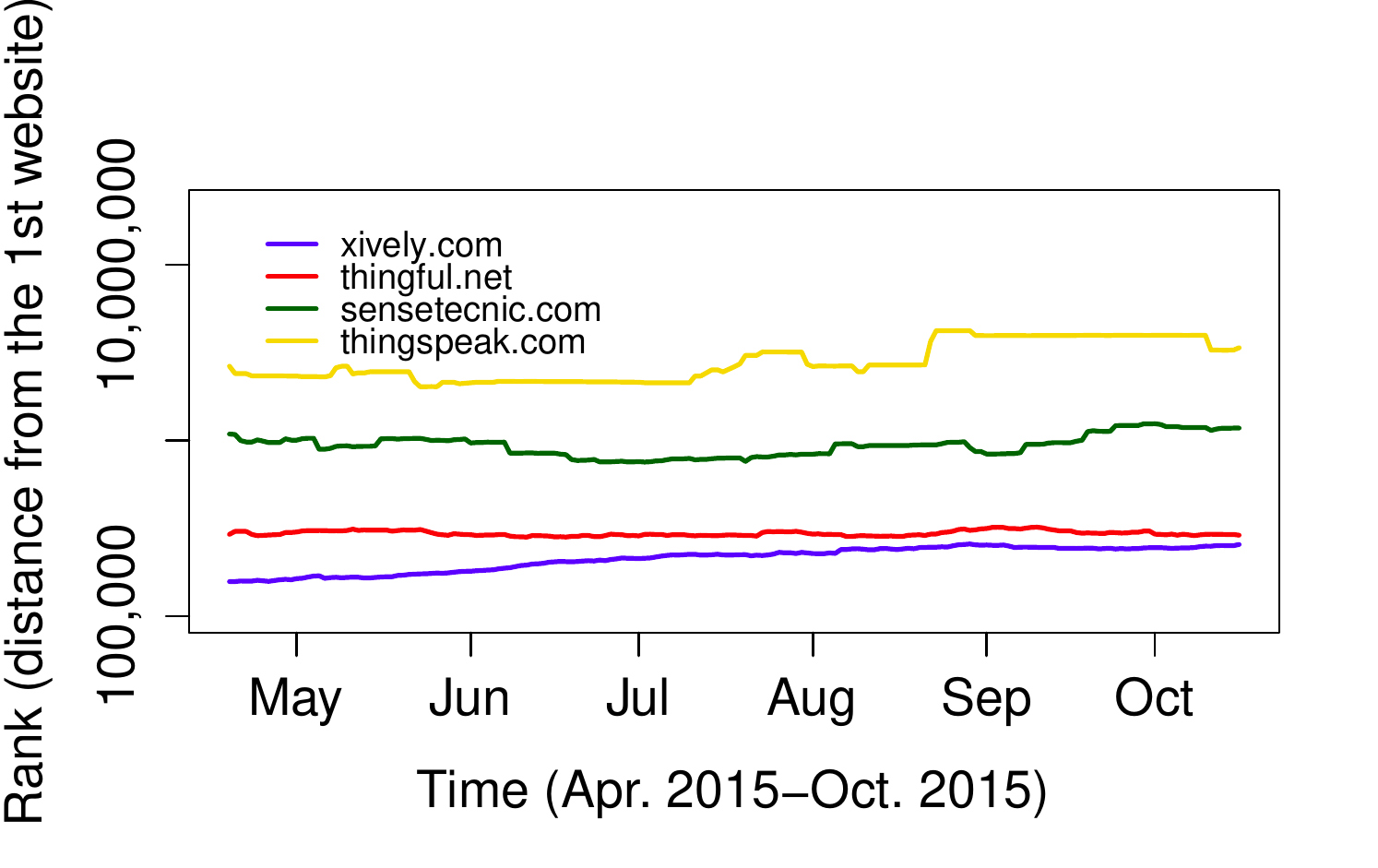}
\caption{Ranking of the popular IoT services}
\label{fig:alexarank}
\vspace{-3mm}
\end{figure}

%
To further understand this trend, we select some of the most cited IoT platforms (i.e., Xively, ThingSpeak, sensetecnic.com, and Thingful) 
from the literature and compare their popularity using Alexa Web Ranking\footnote{http://www.alexa.com}. Figure \ref{fig:alexarank} shows the results for the selected websites during six months from 16 Apr., 2015 to 15 Oct., 2015. 
Accordingly, the popularity of cloud based IoT platforms (e.g., Xively) have been gradually decreasing throughout the last six months while the popularity of Thingful search engine has been increasing during the same period. 
Clearly, a powerful search engine for IoT can help attract users' interests. 

\begin{figure}[!tb]
\centering
\includegraphics[width=.8\linewidth,trim={0 0 0 40},clip]{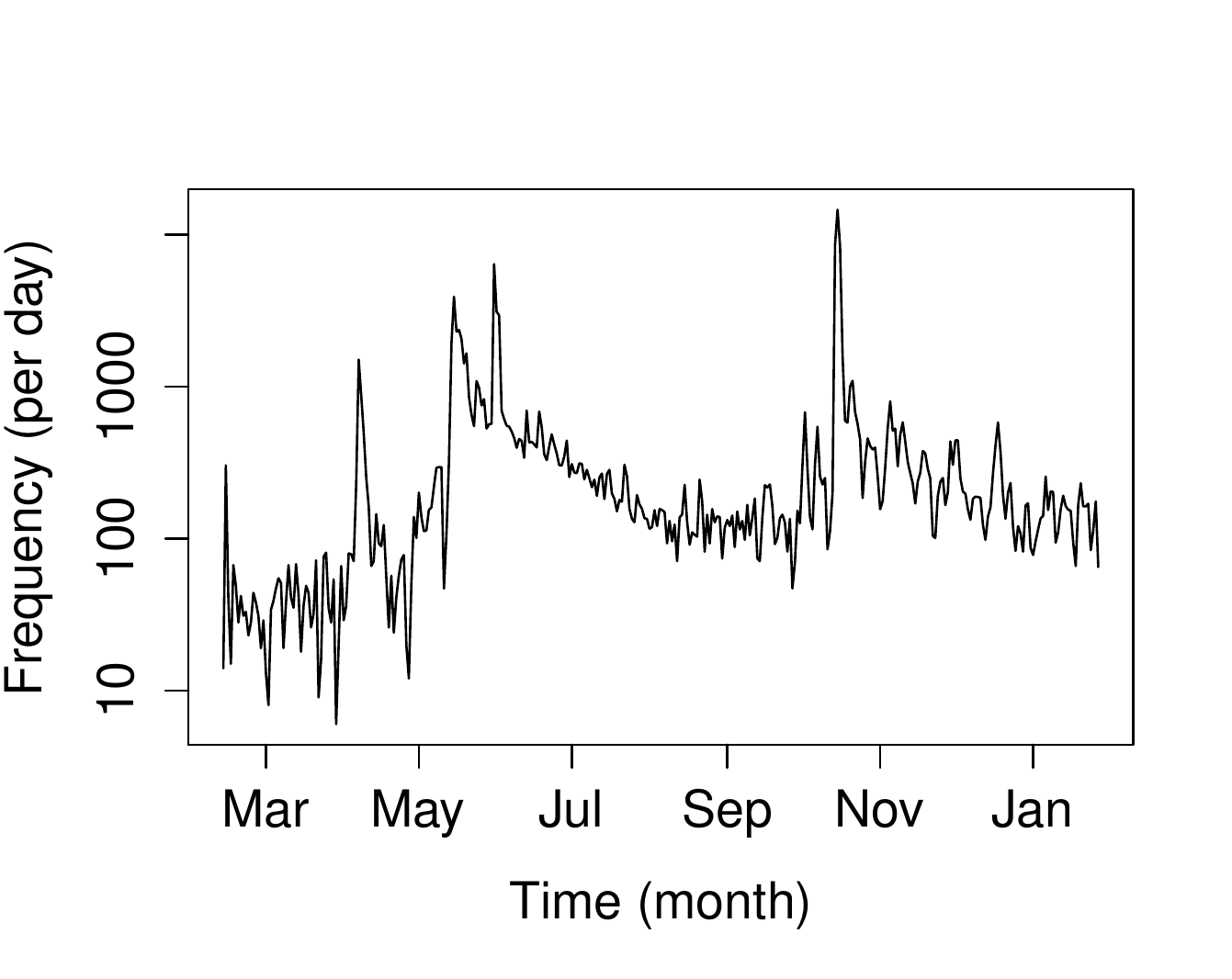}
\vspace{-5mm}
\caption{Query frequency per day in Thingful}
\label{fig:thingfulqt}
\vspace{-3mm}
\end{figure}

\subsubsection{Search Queries Statistics}
Analyzing real-world IoT search queries can provide valuable insights for the design and development of future IoT search engines. To get the statistics, we use a dataset of search queries from the Thingful search engine. Figure \ref{fig:thingfulqt} shows the number of IoT search queries per day. 
It has been gradually increasing through the time and the average number of queries have been tripled since the beginning. 
However, in three points of time, during May, June and October, 2014, an 
abrupt increase in the number of queries per day can be observed. One of the reasons for such increase can be the introduction of new features by the search engine such as embedding and the release of the beta version. This also denotes that any novel improvement in this area can attract many users in a relatively short period of time.


\begin{table}[!tb]
\center
\caption{Most popular keywords and their categories}
\label{tab:thingfulkwd}
\begin{small}
\begin{tabular}{|r|l|r|l|r|}
  \hline
 & \textbf{keyword} & \textbf{freq} & \textbf{category} & \textbf{\% }\\ 
  \hline
  1 & air quality & 71,700 & environment & 61.7\\ 
  \hline
  2 & sensor & 3,348 & misc. & 2.8\\ 
  \hline
  3 & ship & 1,851 & transport & 1.6\\ 
  \hline
  4 & radiation & 1,825 & environment & 1.5\\ 
  \hline
  5 & earthquake & 1,601 & environment & 1.4\\ 
  \hline
  6 & gamma & 1,131 & environment & 1.0\\ 
  \hline
  7 & weather & 876 & environment & 0.8\\ 
  \hline
  8 & shark & 851 & flora and fauna & 0.7\\ 
  \hline
  9 & temperature & 581 & environment & 0.5\\ 
  \hline
  10 & camera & 397 & home & 0.3\\ 
  \hline
  11 & car & 392 & transport & 0.3\\ 
  \hline
  12 & iphone & 271 & home & 0.2 \\ 
  \hline
  13 & fridge & 259 & home & 0.2\\ 
  \hline
  14 & webcam & 255 & home & 0.2\\ 
  \hline
  15 & aircraft & 247 & transport & 0.2\\ 
  \hline
  16 & sharks & 245 & flora and fauna & 0.2\\ 
  \hline
  17 & energy & 242 & energy & 0.2\\ 
  \hline
  18 & food & 239 & home & 0.2\\ 
  \hline
  19 & netatmo & 216 & environment & 0.2 \\ 
  \hline
  20 & coffee & 177 & home & 0.2\\ 
  \hline
  21 & traffic & 168 & transport & 0.1\\ 
  \hline
  22 & transport & 166 & transport & 0.1\\ 
  \hline
  23 & cars & 163 & transport & 0.1\\ 
  \hline
  24 & raspberry pi & 159 & experiment & 0.1\\ 
  \hline
  - & other keywords & 28,771 & - & 24.6 \\
  \hline
  - & Total & 116,131 & - & 100\\ 
  \hline
\end{tabular}
\end{small}
\end{table}

According to the query logs, 84.9\% of queries are associated with keywords. 
An investigation over the popular keywords yields Table \ref{tab:thingfulkwd}. The category is selected from Thingful's predefined categories including Energy, Home, Health, Environment, Flora and Fauna, Transport, Experiment, Miscellaneous. Apparently, environmental sensing related keywords such as ``air quality" and ``radiation" have been very popular amongst users.

The category analysis in Table \ref{tab:thingfulkwd} shows that for the majority of the queries,
transportation related keywords constitute less than 3\%  of the search queries. On the other hand, keywords that are related to the environmental scanning, constitute more than 67\% of the search queries. Thus, in assigning computing resources, environmental data sources should receive more attention. That is, more effort is needed to make the environmental data sources updated and in this way, we can 
use our computing resources more efficiently.

\begin{figure*}[!tb]
\begin{center}
  \subfigure [6pm ACST]
    {\includegraphics[width=0.2\linewidth]
    {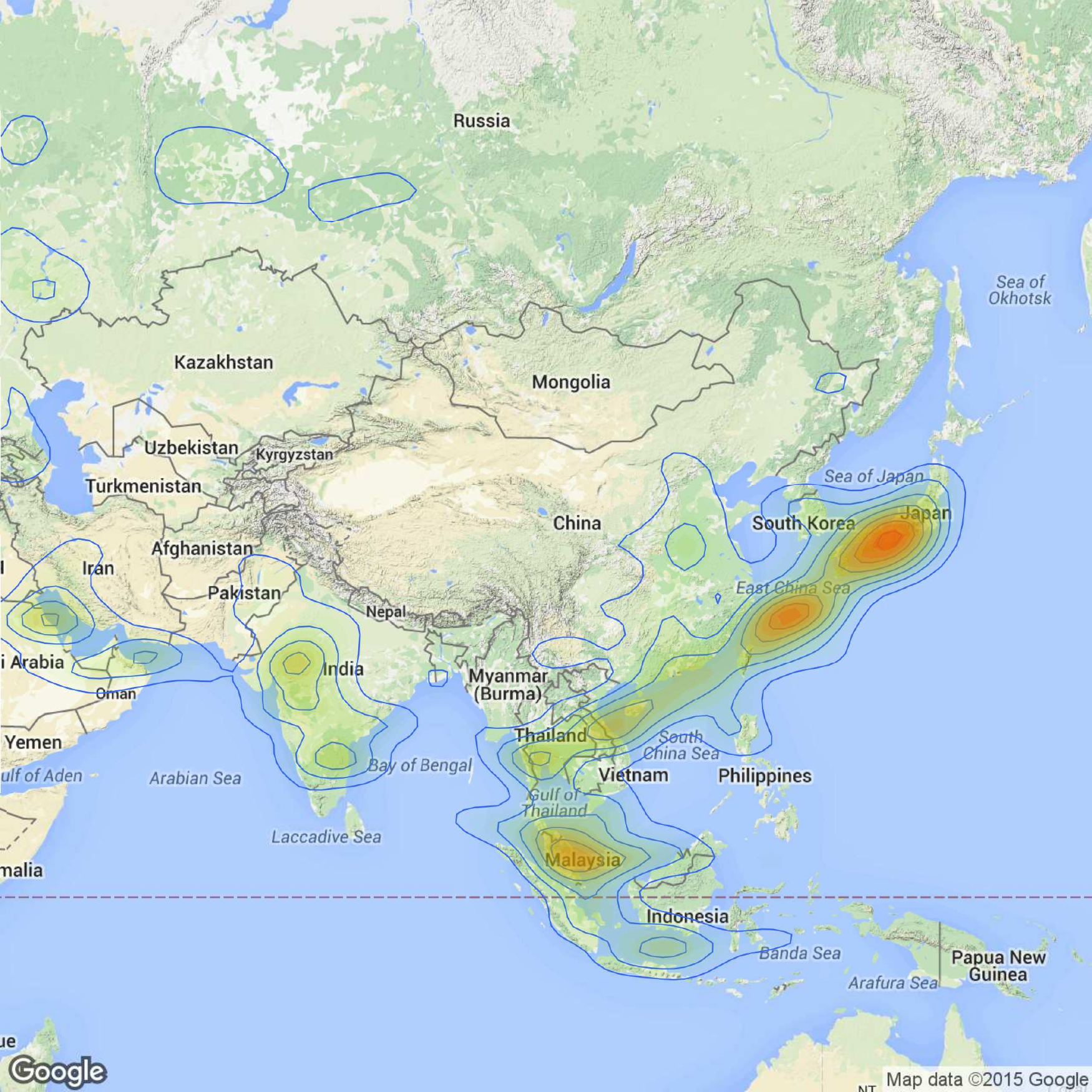}
    \label{fig:asia1}}
  \subfigure [12am ACST]
    {\includegraphics[width=0.2\linewidth]
    {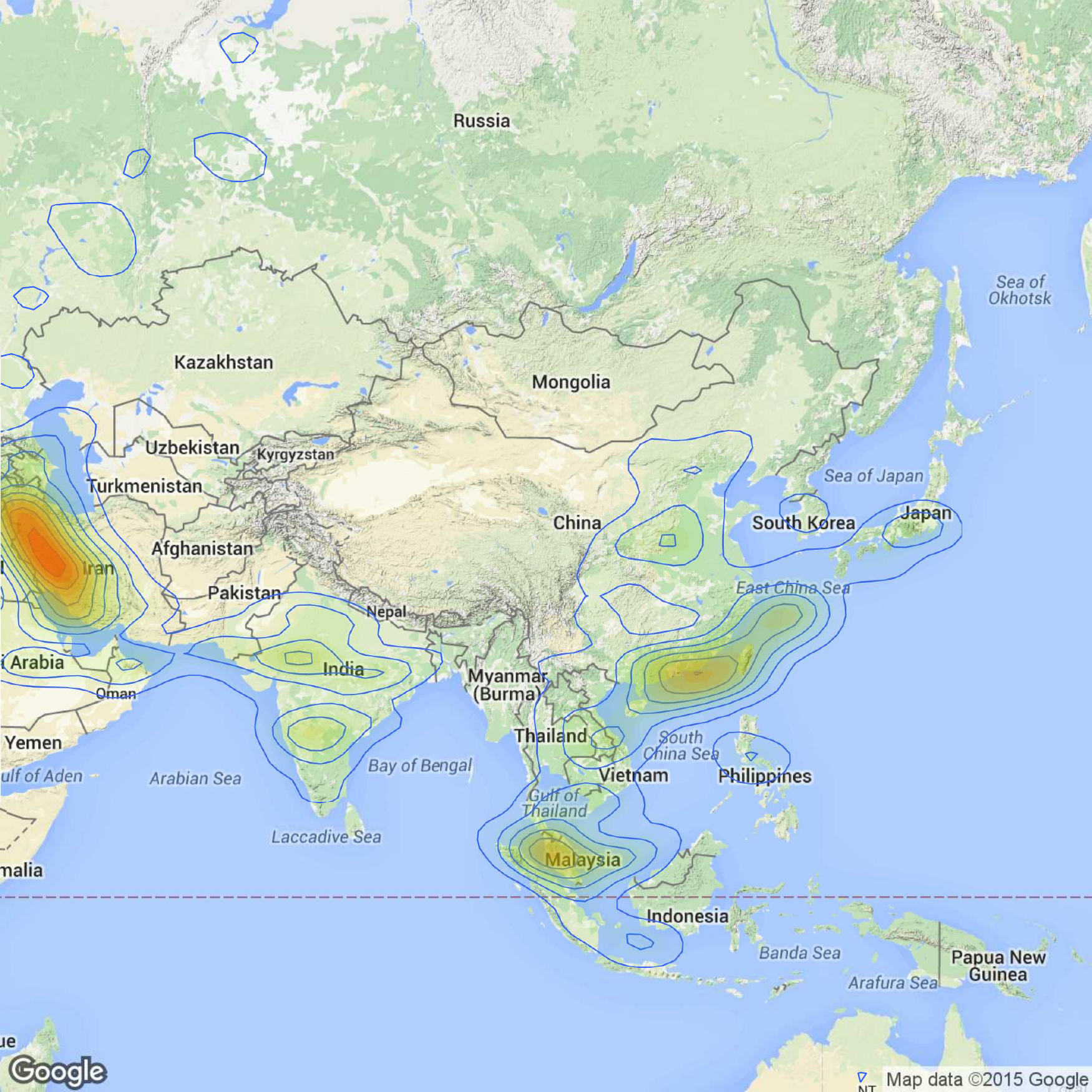}\label{fig:asia2}}
  \subfigure [6am ACST]
    {\includegraphics[width=0.2\linewidth]
    {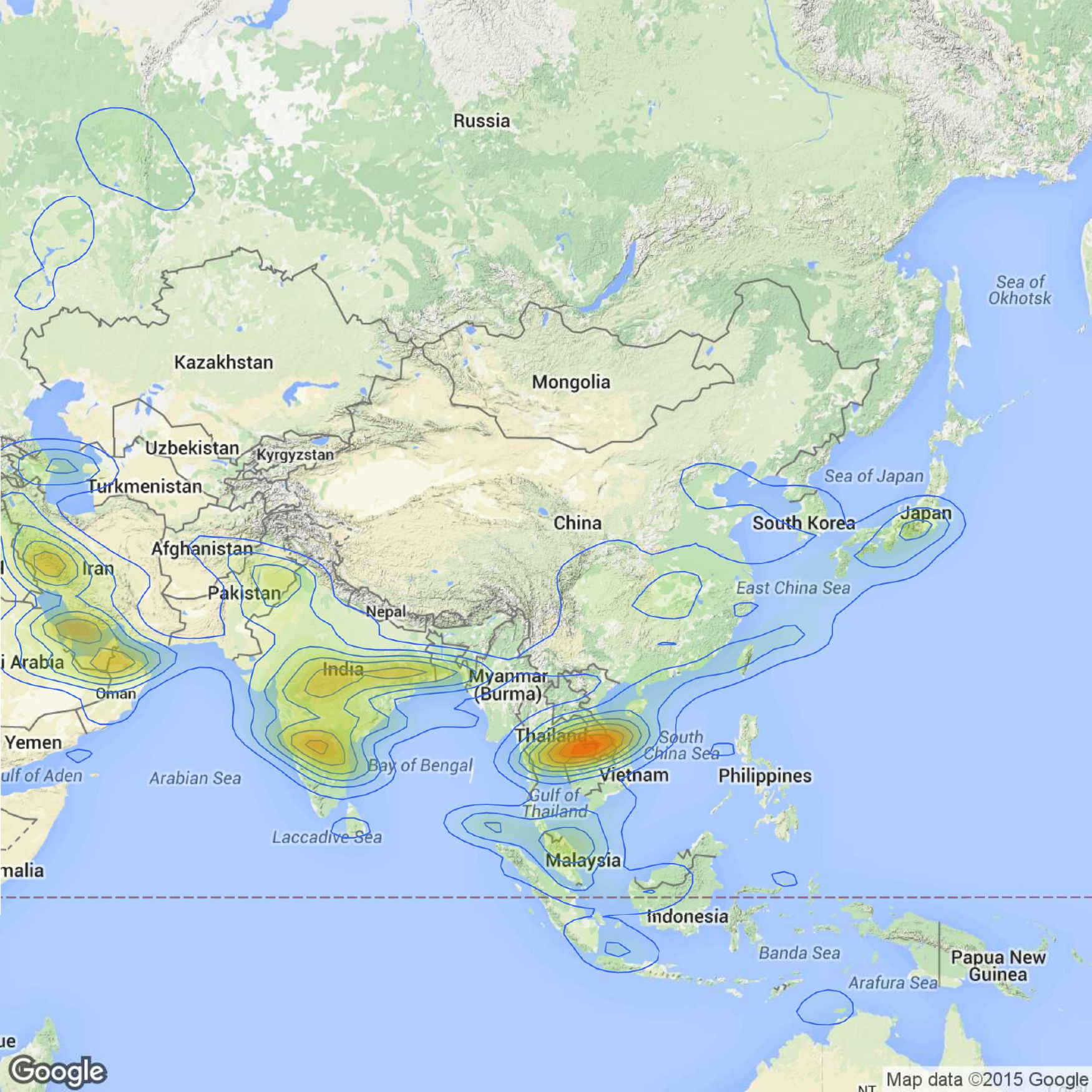}\label{fig:asia3}}
  \subfigure [all the time]
    {\includegraphics[width=0.2\linewidth]
    {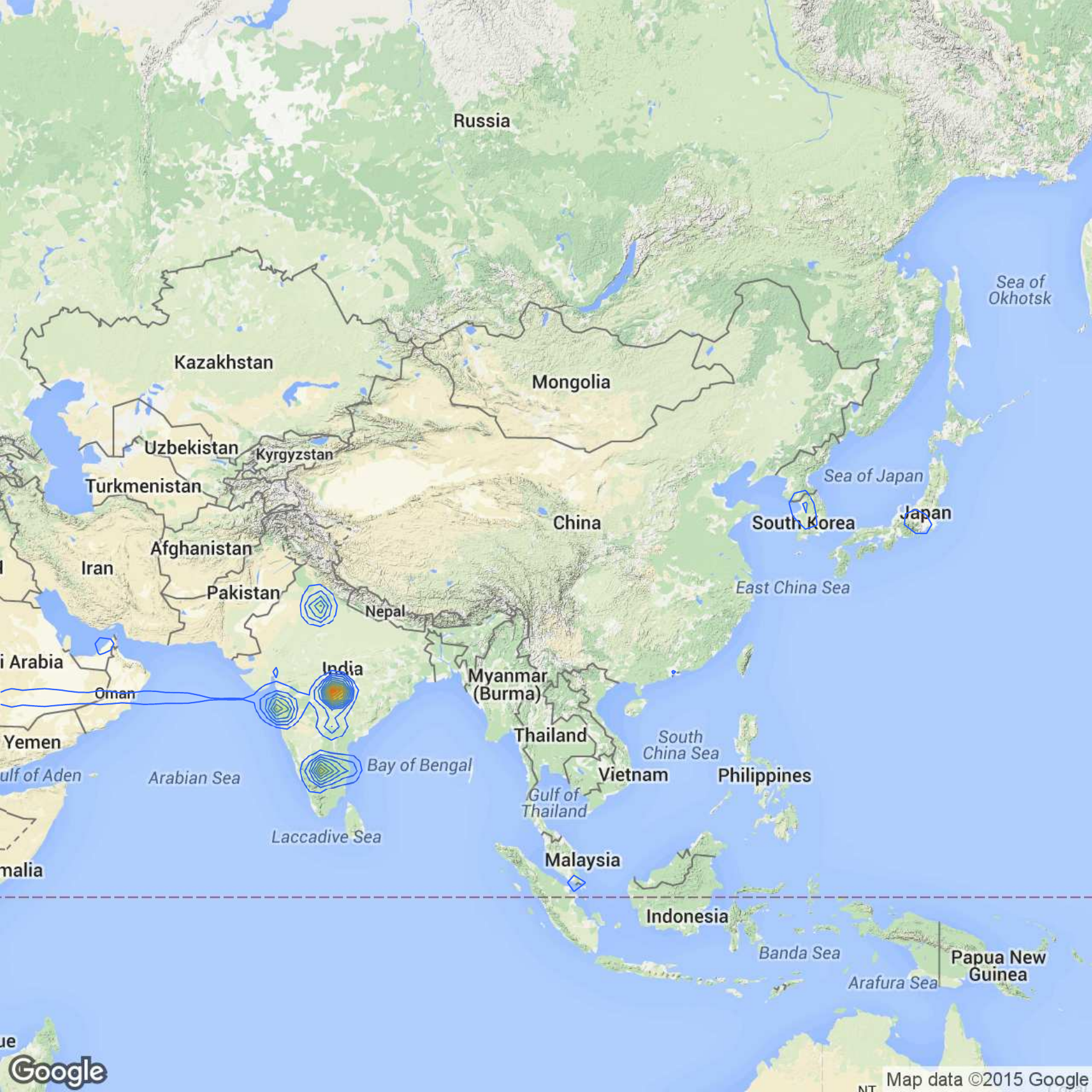}\label{fig:asiaqmap}}
  \subfigure [6pm ACST]
    {\includegraphics[width=0.2\linewidth]
    {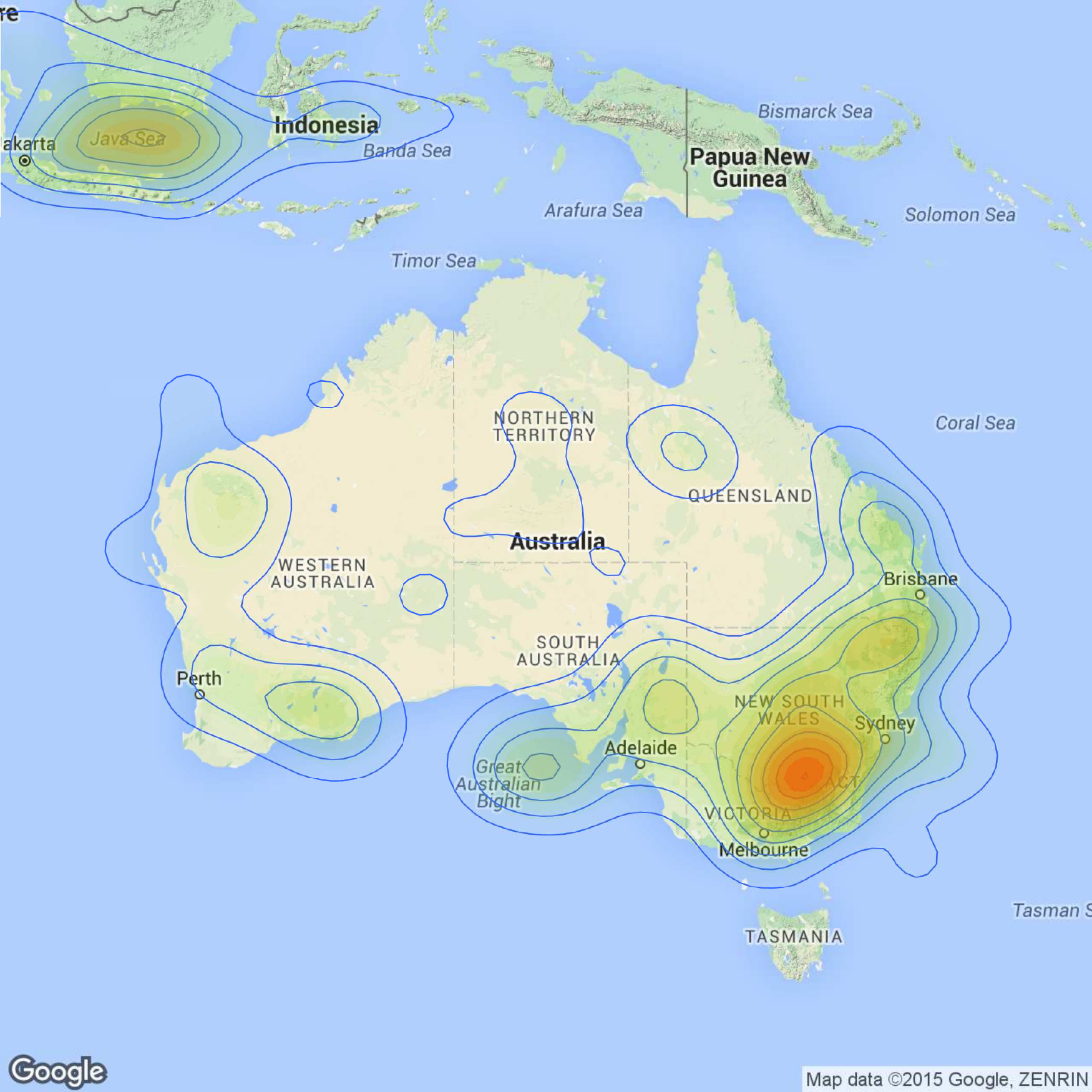}\label{fig:aus1}}
  \subfigure [12am ACST]
    {\includegraphics[width=0.2\linewidth]
    {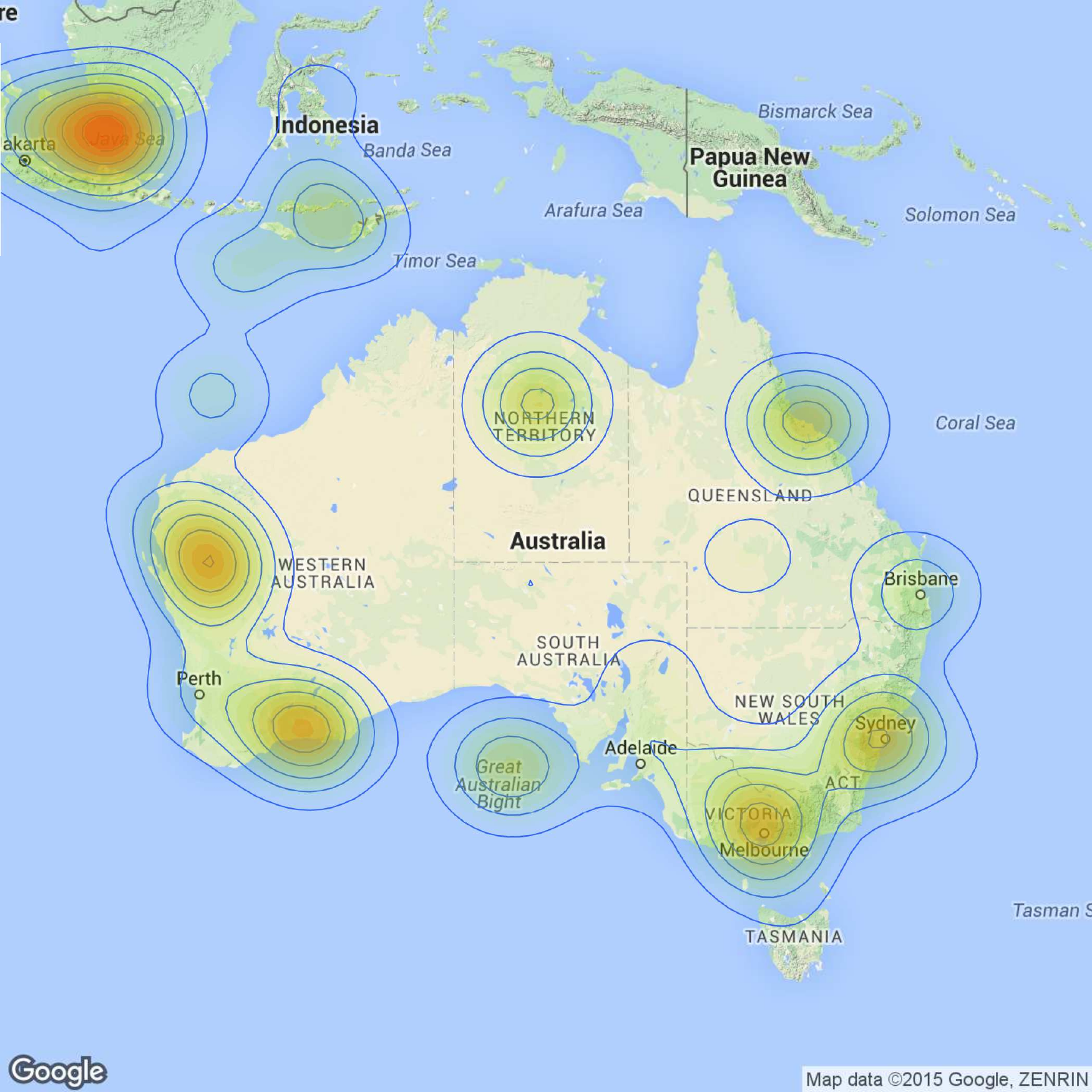}\label{fig:aus2}}
  \subfigure [6am ACST]
    {\includegraphics[width=0.2\linewidth]
    {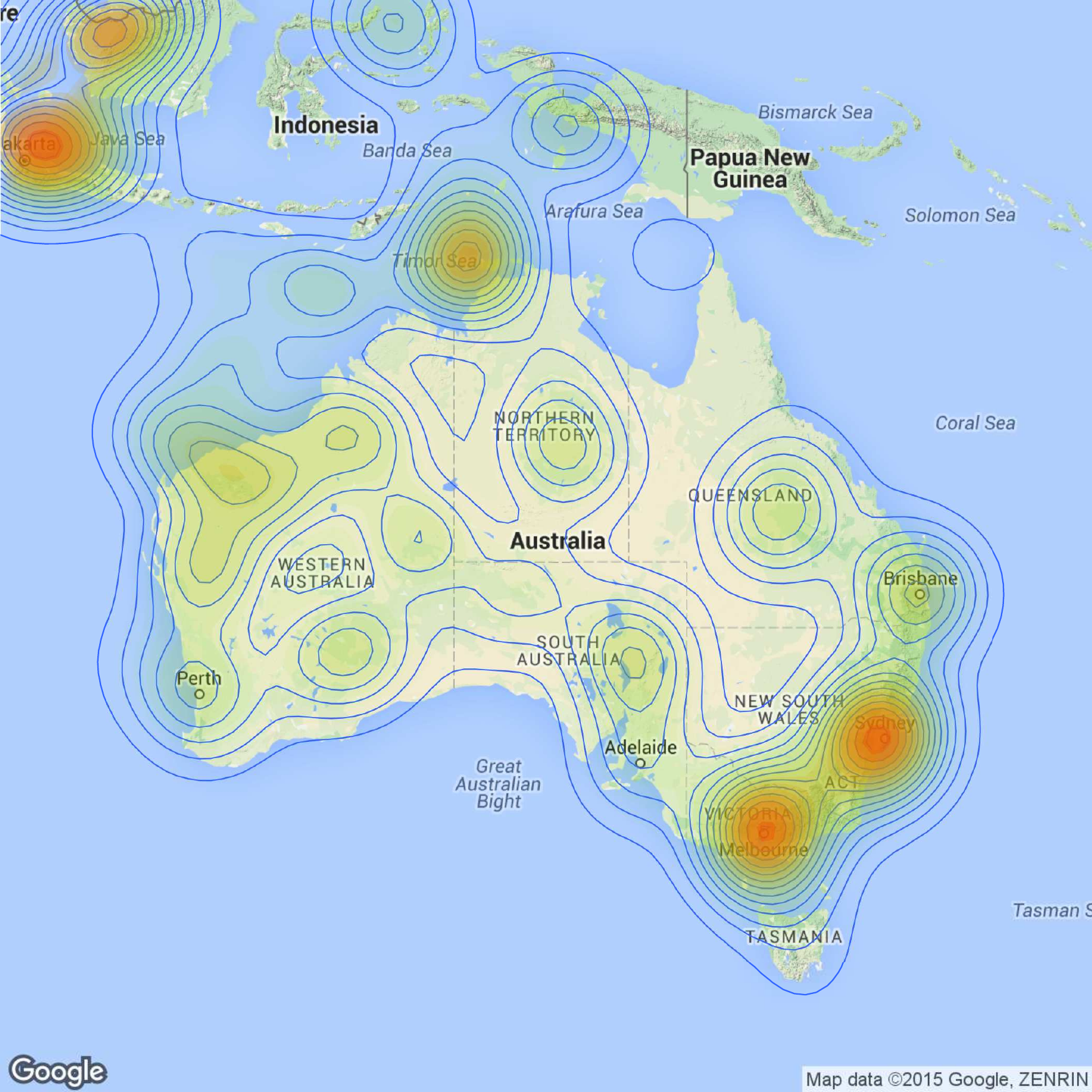}\label{fig:aus3}}  
  \subfigure [all the time]
    {\includegraphics[width=0.2\linewidth]
    {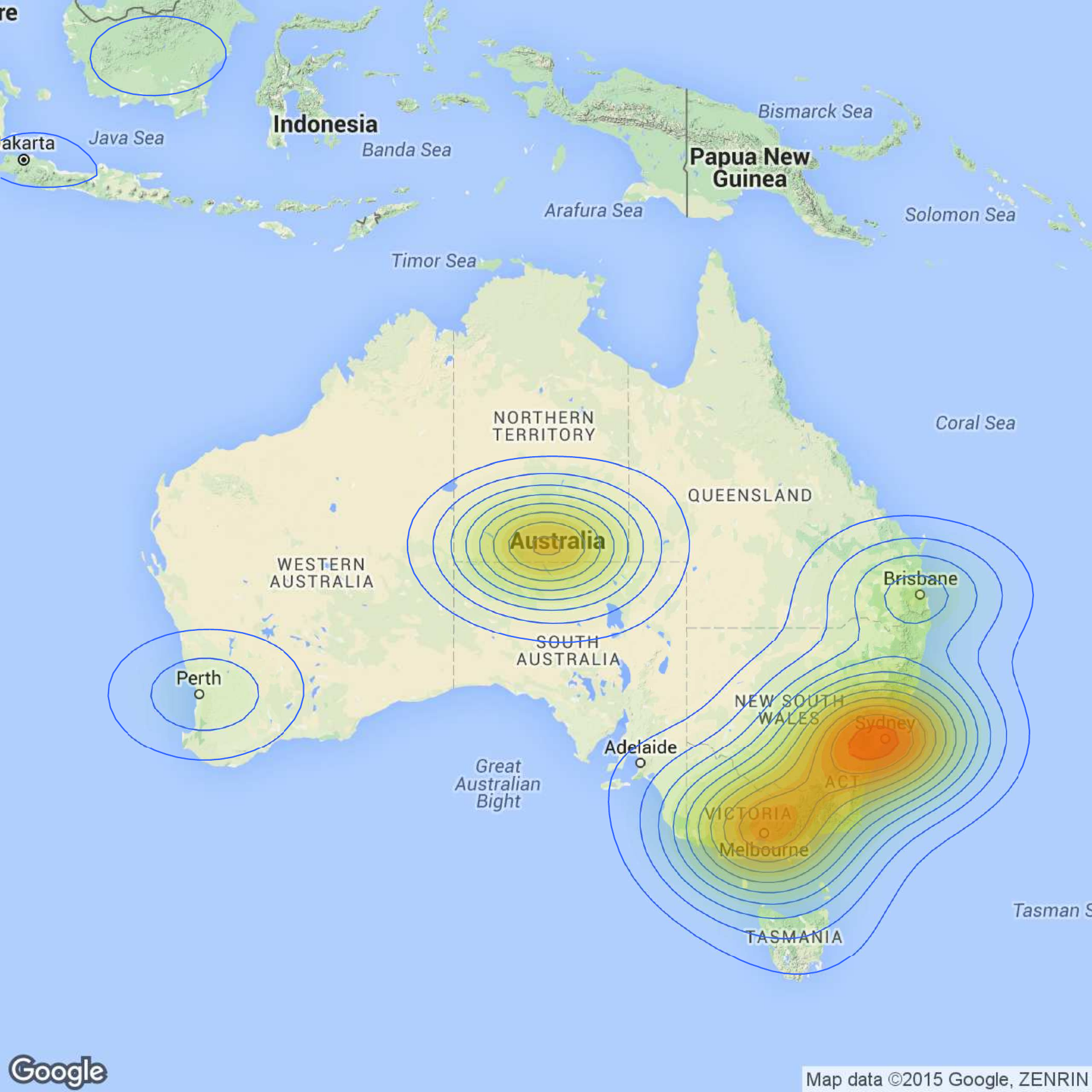}\label{fig:ausqmap}}
  \subfigure [6pm ACST]
    {\includegraphics[width=0.2\linewidth]
    {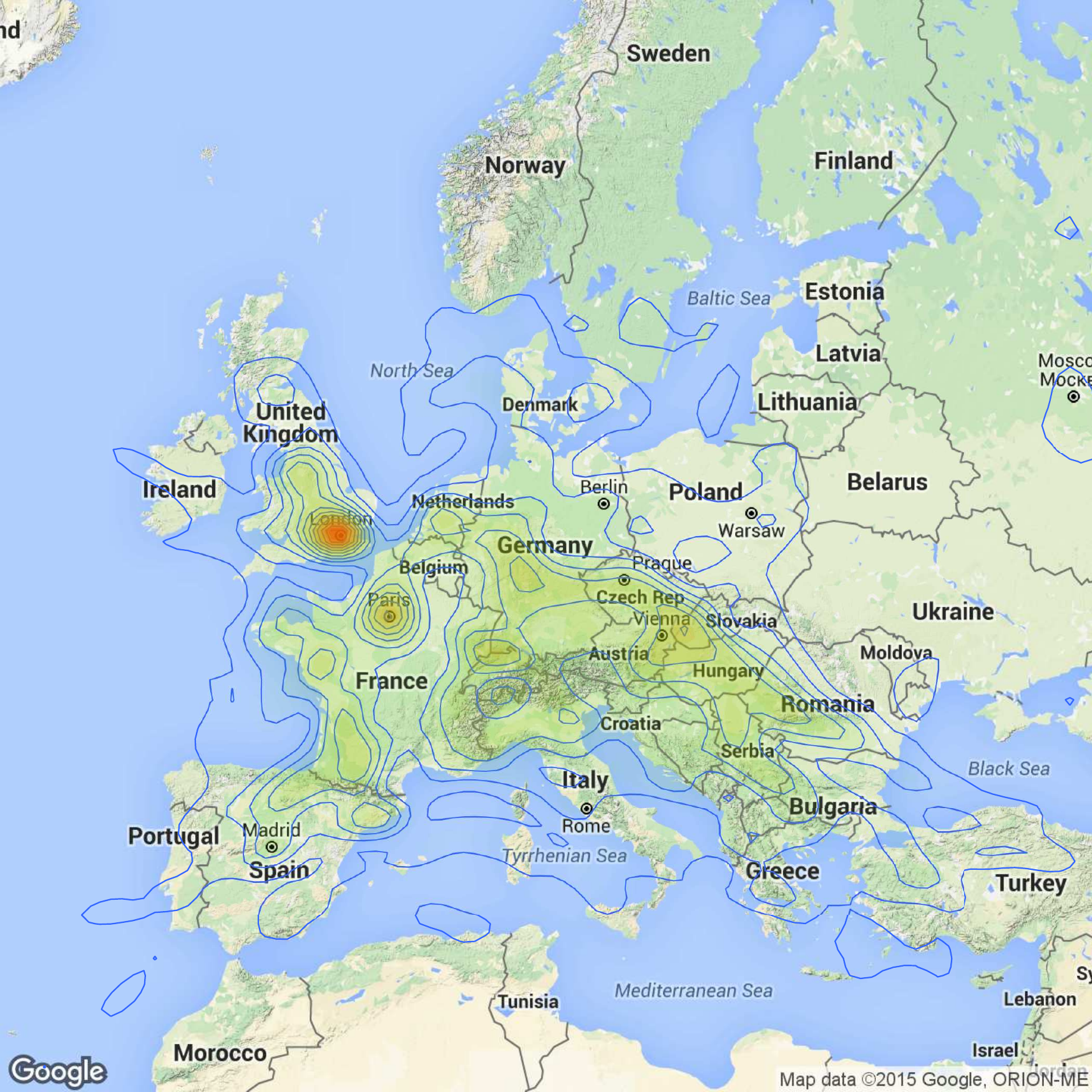}\label{fig:europe1}}
  \subfigure [12am ACST]
    {\includegraphics[width=0.2\linewidth]
    {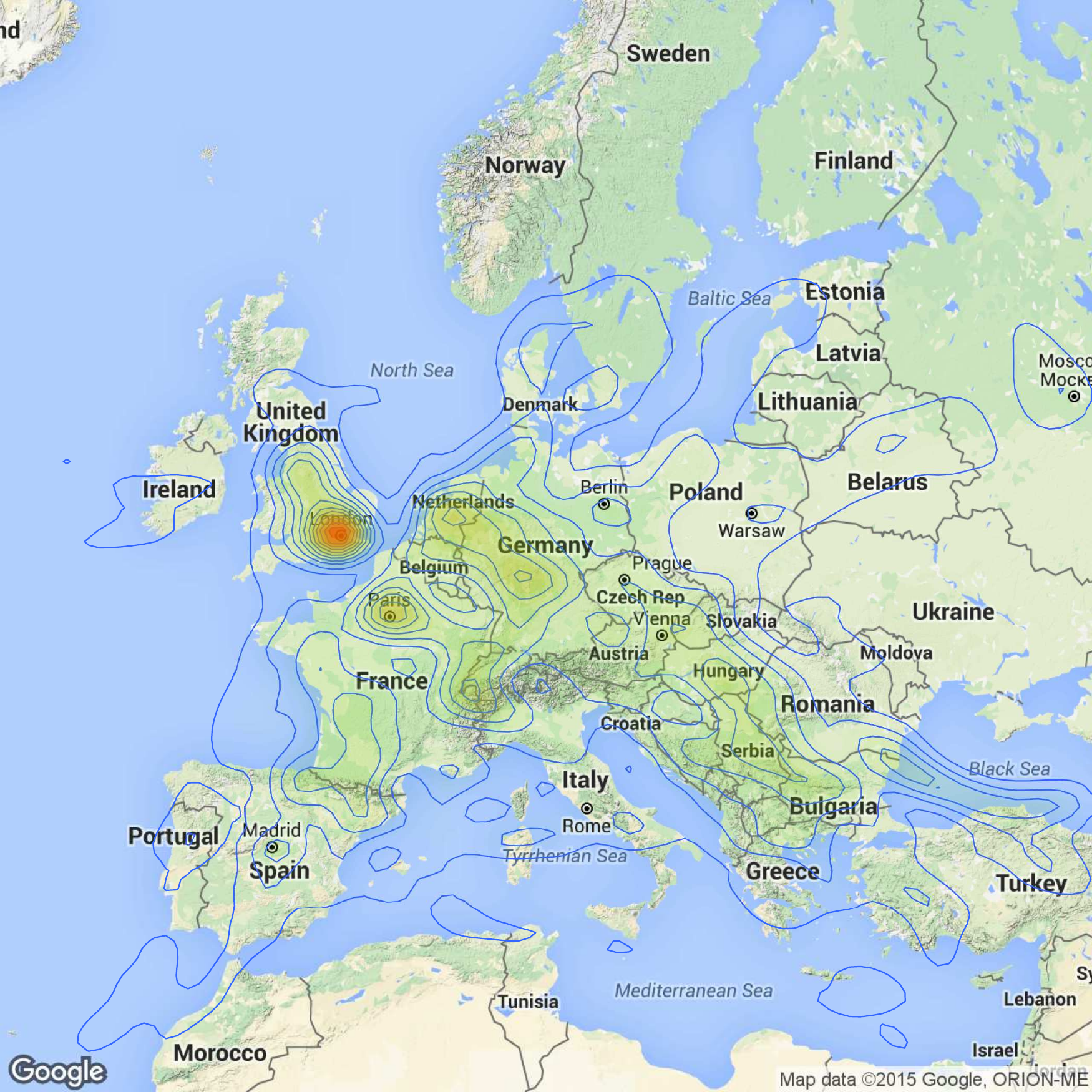}\label{fig:europe2}}
  \subfigure [6am ACST]
    {\includegraphics[width=0.2\linewidth]
    {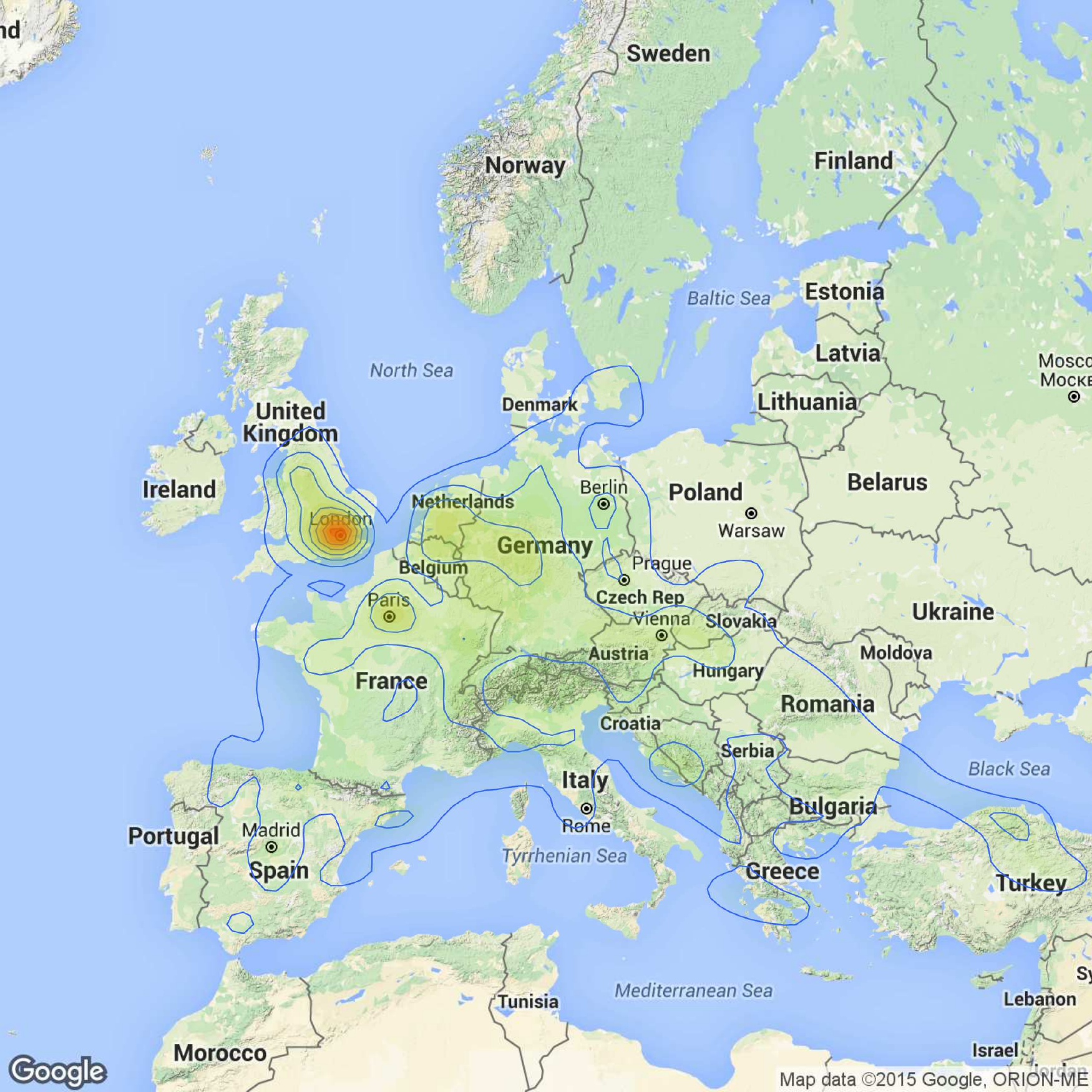}\label{fig:europe3}}
  \subfigure [all the time]
    {\includegraphics[width=0.2\linewidth]
    {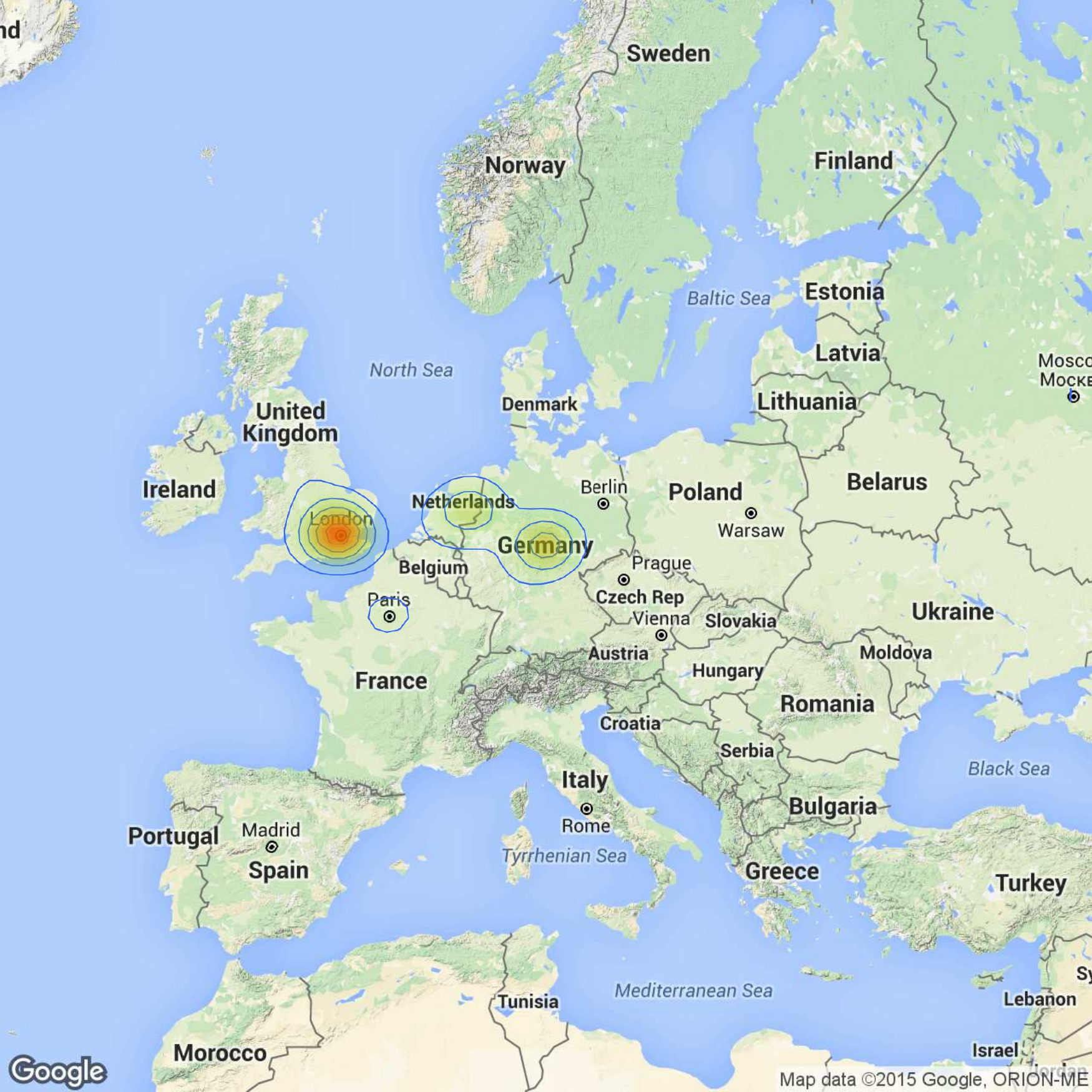}\label{fig:euqmap}}
\caption{
Heat map for things (a,b,c,e,f,g,i,j,k) and user queries (d,h,l)
}
\label{fig:geodist}
\end{center}
\vspace{-3mm}
\end{figure*}

\subsection{IoT Data Characteristics}
IoT data is semi-structured as the popular format in IoT data transmission is JSON. To provide a more detailed vision over IoT data characteristics, we investigate \textit{data source types} and the \textit{dynamics} and the \textit{quality} of IoT data.

\subsubsection{Data Source Types}
\begin{figure}[!tb]
\centering
\subfigure []
    {\includegraphics[width=.49\linewidth]
    {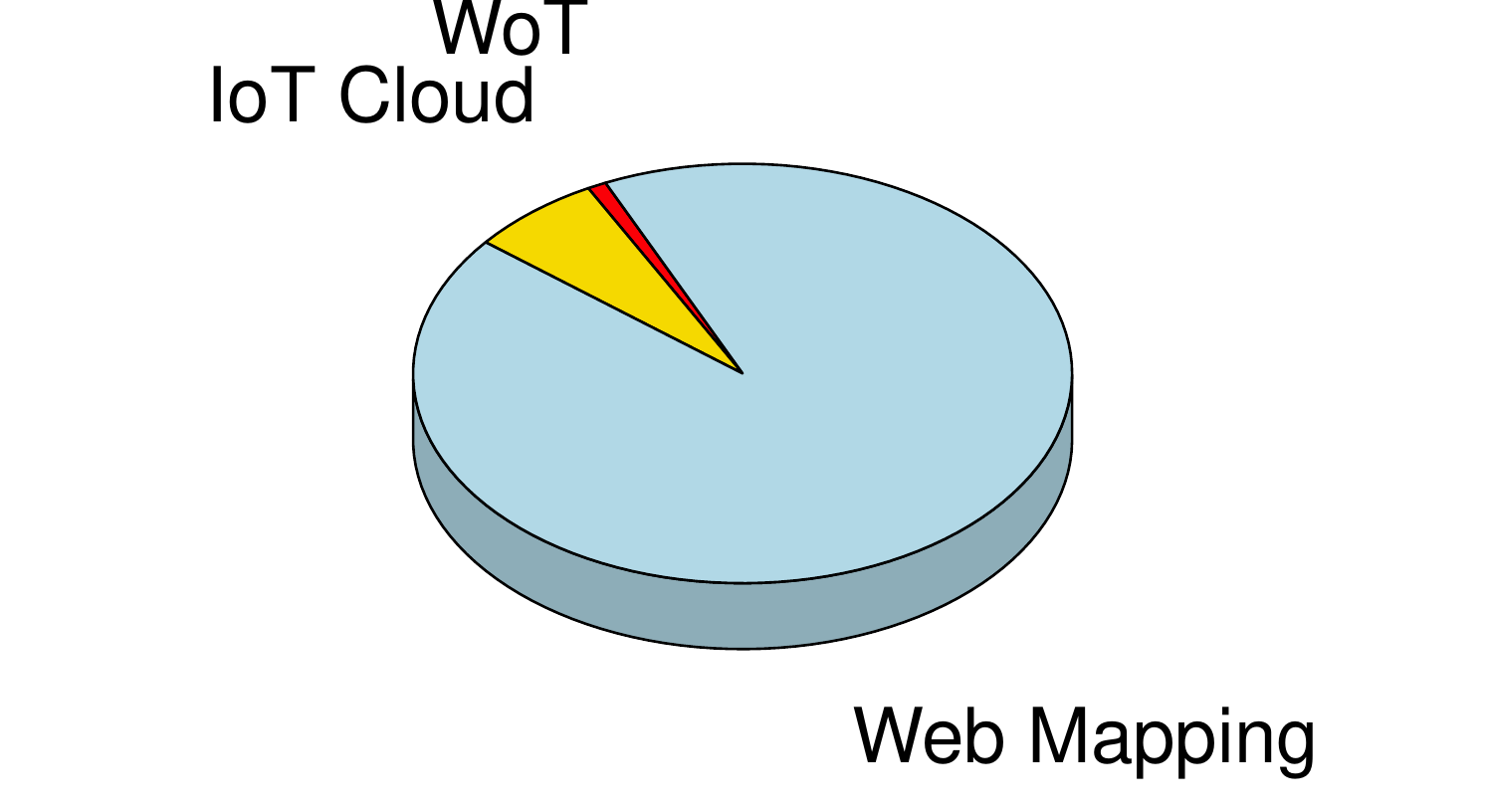}\label{fig:pieiot}}
\subfigure []
    {\includegraphics[width=.49\linewidth]
    {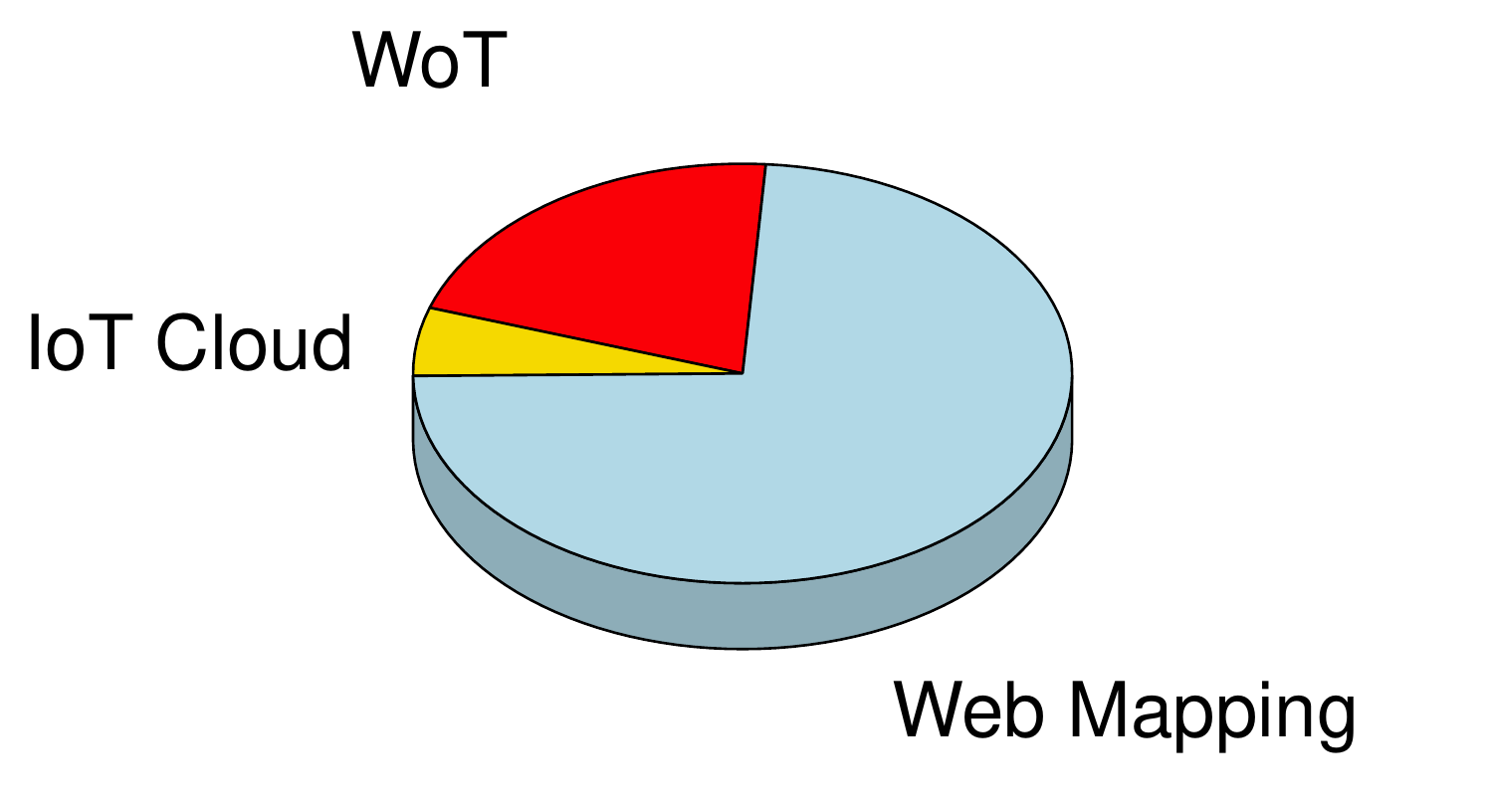}\label{fig:pieiotsites}}
    \caption{Major provider technologies for public IoT data based on the (a) number of things; and (b) number of websites}
    \vspace{-2mm}
\end{figure}

As Figure \ref{fig:pieiot} suggests, amongst the publicly available data sources, Web Mapping serves the majority of things (88\%), which is followed by IoT Cloud services (7\%) and WoT (1\%). However, in exchange we observe that a larger number of data sources use WoT to share IoT data on the Web (Figure \ref{fig:pieiotsites}) while most of the IoT cloud platforms do not provide access to any public data.

To grasp a more detailed image of IoT clouds and WoT, Table \ref{tab:wotcloud} shows WoT and IoT clouds and the number of non-private things which use these technologies.

\begin{table}[!tb]
\caption{WoT vs. IoT cloud services}
\label{tab:wotcloud}
\centering
\begin{small}
    \begin{tabular}{|p{0.3\linewidth}|p{0.3\linewidth}|l|}
    \hline
    \textbf{Data Source}                       & \textbf{Public Sensors (Things)} & \textbf{Type}                                    \\ \hline
    Xively                              & \~ 67,000                & IoT Cloud                             \\ \hline
    WoTkit                              & 4,065                    & WoT    \\ \hline
    ThingSpeak                          & 3,571                    & WoT\\ \hline
    WikiBeacon                          & 30,052                   & WoT              \\ \hline
    ISMN* & 2,080                    & WoT    \\ \hline
\multicolumn{3}{l}{*International Soil Moisture Network}
    \end{tabular}
\end{small}
\end{table}

\subsubsection{Spatial and Temporal Distribution of Things}
Understanding the spatial and temporal distribution of IoT and query updates is valuable for identifying the existing gaps, which can help in predicting the trends of searches and updates. To model the spatial gaps between IoT and queries, we use the Earth Mover's Distance (EMD) measure. EMD describes the normalized minimum amount of work required to transform one distribution to the other. In our case, given the two distributions matrices of things, $d_1$ and $d_i$, which have been taken in timestamps $t_1$ and $t_i$ respectively, we want to measure the amount of changes in the latter distribution ($d_i$) from the initial distribution $d_1$ using $EMD(d_1,d_i)$ measure. Therefore, we can monitor the changes in distribution over the time.



\begin{figure}[!tb]
\centering
\vspace{-2mm}
\includegraphics[width=.7\linewidth,trim={0 1 0 30}, clip]{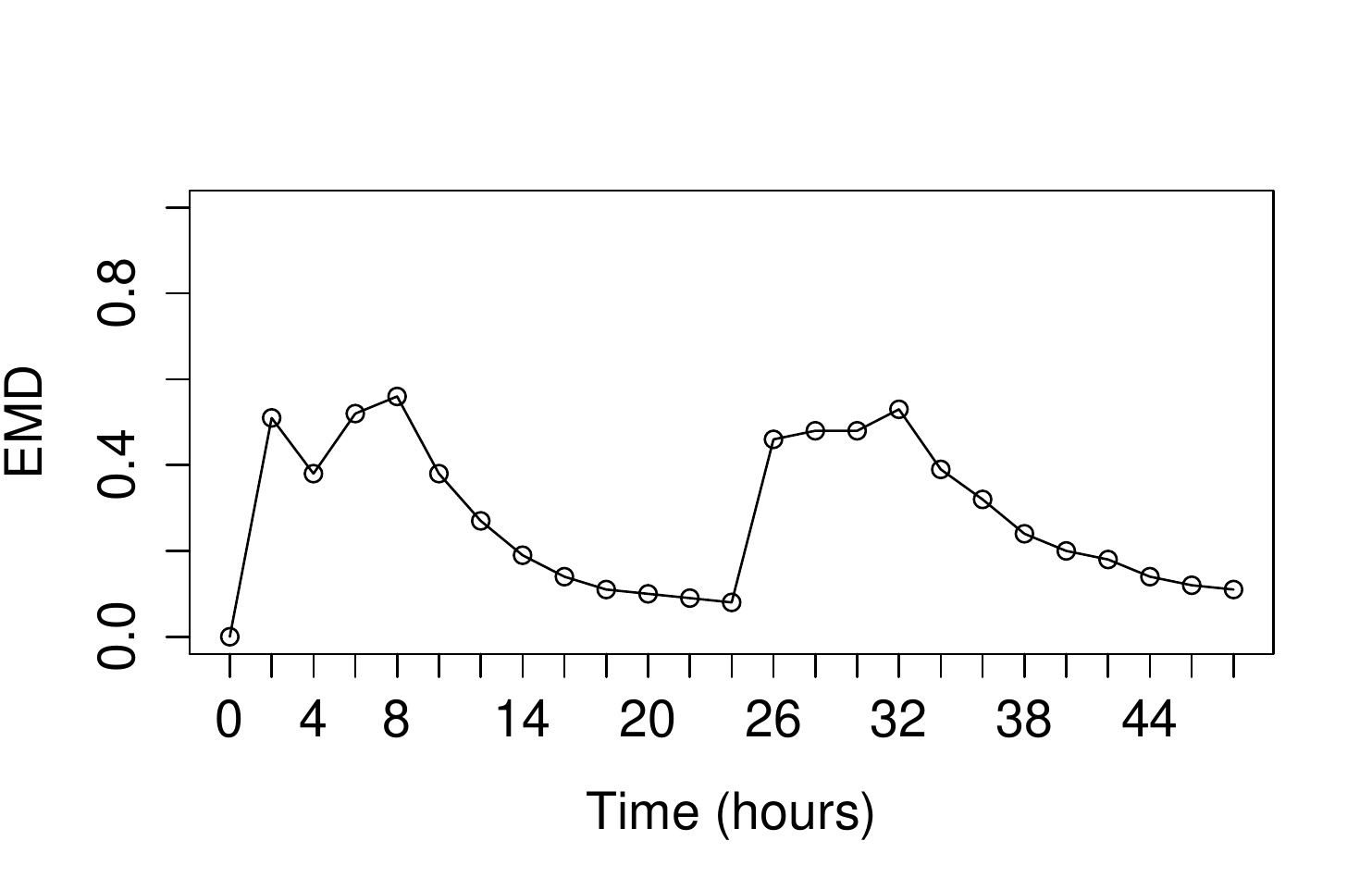}
\caption{EMD score for things data during 48 hours}
\label{fig:densitytime}
\vspace{-3mm}
\end{figure}

In the next step, we want to know whether the patterns in Figure \ref{fig:geodist} and changes in the distribution of things recur over the time. 
Thus, we perform the same analysis over a period of time on how the spatial distribution of things changes through the time.
In particular, we use the emdist \cite{emdist2015} implementation to approximate the EMD score for each transition.
Figure \ref{fig:densitytime} shows the EMD score for a given period of time in 48 timetamps 
which we have collected within 48 hours. The curve shows that in each given timestamp $t_i$, the value of $EMD(t_1,t_i) \in [0,1]$. Thus, the EMD score for $t_1$ is 0 
since there is no difference between the distribution matrix in $t_1$ and itself. Shortly, a huge amount of change is observed between $t_2$ to $t_5$ and later the EMD score continuously decreases as the distribution returns to its initial status. The very same pattern recurs on the next period of time. As a result, we understand that the geospatial distribution of things goes back to its original state over a period of time (in this case, after 24 hours). This result can assist in setting up new strategies for saving computing resources when updating the things dataset. For example, during an update process, we can scan the areas with higher densities more often than the lesser dense areas.

\begin{table*}[]
\center
\begin{small}
\caption{Example of selected parameters from a set of readings of a specific sensor in Xively}
\label{tab:xivelyrec}
\begin{tabular}{|l|l|l|l|l|l|l|l|l|}
\hline
\multicolumn{1}{|c|}{\textbf{}} & \multicolumn{1}{c|}{\textbf{id}} & \multicolumn{1}{c|}{\textbf{title}} & \multicolumn{1}{c|}{\textbf{private}} & \multicolumn{1}{c|}{\textbf{status}} & \multicolumn{1}{c|}{\textbf{updated}} & \multicolumn{1}{c|}{\textbf{Created}} & \multicolumn{1}{c|}{\textbf{value}} & \multicolumn{1}{c|}{\textbf{Symbol}} \\ \hline
1                               & 1213                             & house                               & false                                 & public                               & 2015-06-10T13:01:59.997058Z           & 2008-11-27T18:20:25.169483Z           & 77.46                               & F                                    \\ \hline
2                               & 1213                             & house                               & false                                 & public                               & 2015-06-10T20:31:00.777699Z           & 2008-11-27T18:20:25.169483Z           & 78.56                               & F                                    \\ \hline
3                               & 1213                             & house                               & false                                 & public                               & 2015-06-11T03:50:00.136106Z           & 2008-11-27T18:20:25.169483Z           & 79.5                                & F                                    \\ \hline
\end{tabular}
\end{small}
\vspace*{-2mm}
\end{table*}

\subsubsection{Data Dynamics}
\begin{figure}[tb]
\begin{center}
\includegraphics[width=0.7\linewidth,trim={0 15 0 50}, clip]{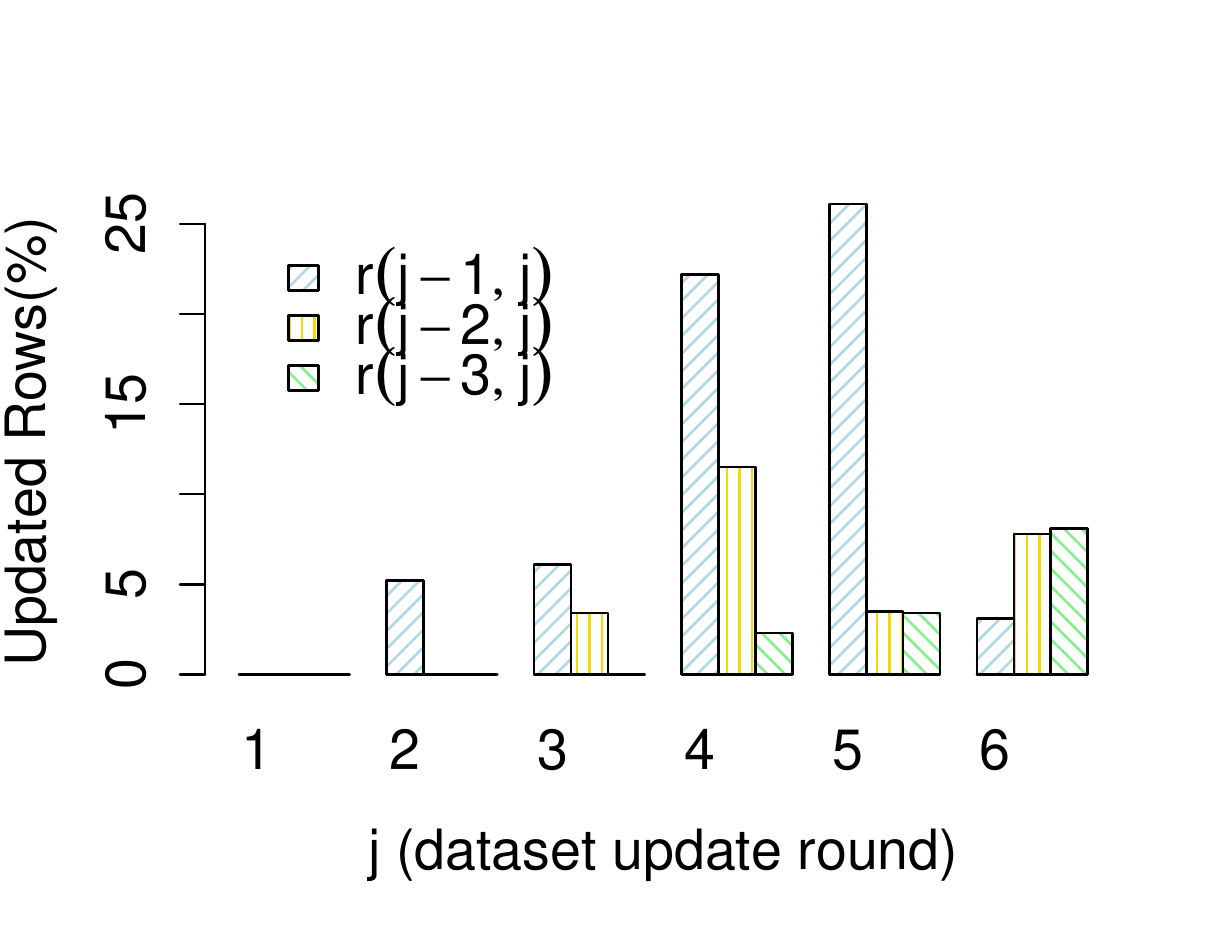}
\vspace{-2mm}
\caption{Ratio of the rows with new sensor outputs}
\label{fig:xivelyupds}
\end{center}
\vspace{-4mm}
\end{figure}

IoT data are widely regarded as highly dynamic and volatile \cite{qinSFDWV2014}. Although several approaches have been proposed to tackle various problems caused by the dynamic nature of IoT, no other work investigates the real-world IoT data on their dynamics. With our first-hand dataset collected, we observe the following interesting aspects on IoT data:
\begin{itemize}
\item Only a small portion of IoT data changes frequently. 
This finding can be easily checked by measuring the number of things and the amount of data that is being updated (new sensor reading during the next IoT scan). 
For instance, Figure \ref{fig:xivelyupds} shows the ratio of things which have updated their previous readings from nearly 70,000 objects on the Xively network, which is a part of our things dataset. The ratio of updated rows $r \in [0,1]$, is obtained from the following equation:
\begin{equation}
r(i,j) = \frac{|diff(d_i,d_j)|}{max(|d_i|,|d_j|)}
\end{equation}
where if $D$ is the domain of sensor readings, $d_i \in D$ denotes sensor readings in timestamp $i$, $diff:D\times D \rightarrow N^+$ is a function which returns the new rows in $d_j$ and also $j > i$. 
Here, the time difference between each $j-1$ and $j$ is 6 hours.
As shown, 
up to 23\% of objects have new sensor outputs during the experiment. 
Furthermore, only a small part of each tuple gets updated each time. Table \ref{tab:xivelyrec} shows an example from the Xively platform. We only select a small subset of the attributes (77 attributes in the original version) for the illustration purpose. 
As it shows, only the \textit{value} attributed is being updated every time. 

\item Frequencies of updates for the same object from different data sources can be highly variant. 
For instance, with every flight tracker website the sensor readings for flights are updated several times per second, while in MarineTraffic the sensor readings for ships and vessels are updated 
every three minutes. 

\item Similar to the geographical distribution of objects, IoT dynamics may follow patterns over the time. As Figure \ref{fig:xivelyupds} shows, the ratio of updated values decreases when increasing the number of steps. 
This indicates that many of the updated tuples return to their initial values after a while. For example, an air quality egg, which is an egg shaped device to measure indoor temperature and air quality, may report the similar temperature in 24 hours.
\end{itemize}

\subsubsection{Data Quality}

We observe that different data sources, may share the data that is being generated by the same sensors. 
One of the interesting points in the integration of IoT data would be knowing the redundancy. Also consistency of the redundant data will be an interesting topic for researchers.

Table \ref{tab:duplicatesites} shows a list of redundant sources of IoT data and the type of the things that they cover. We select a few data sources which seem to be more popular from three different categories: \textit{flight tracking} and \textit{marine traffic tracking}. Every object from these sources is associated with an identifier which can distinguish it from other objects. We merge the data from all websites in each category to get the ratio of inclusiveness. This measure denotes the rate of the objects which exist in the data source and the union set of objects for the corresponding category.

\begin{table*}[!tb]
\begin{small}
\begin{center}
\caption{Transportation data sources with overlapping set of objects}
\label{tab:duplicatesites}
\begin{tabular}{|l|l|l|c|c|c|}
\hline
\multicolumn{1}{|c|}{\textbf{Application}}                                             & \textbf{URL}                                 & \multicolumn{1}{c|}{\textbf{Scope}} & \textbf{Novel Data} & \textbf{Inclusiveness} & \textbf{Delay} \\ \hline
\multirow{7}{*}{\begin{tabular}[c]{@{}l@{}}Flight \\ tracking\end{tabular}}            & http://www.flightradar24.com                 & \multirow{5}{*}{Worldwide}          & \checkmark                 &                       99.99\%  & -              \\ \cline{2-2} \cline{4-6} 
                                                                                       & https://flightaware.com/live/                &                                     & \checkmark                 &        0.01\%                & \checkmark              \\ \cline{2-2} \cline{4-6} 
                                                                                       & https://planefinder.net                      &      & \checkmark                 &             0.01\%       & \checkmark              \\ \cline{2-2} \cline{4-6} 
                                                                                       & http://www.radarbox24.com                    &                                     & \checkmark                 &                0.01\%        & \checkmark              \\ \cline{2-2} \cline{4-6} 
                                                                                       & http://www.radarvirtuel.com                  &                                     & \checkmark                 &              0.01\%          & \checkmark              \\ \cline{2-6} 
                                                                                       & http://tinyurl.com/klmliv                    & Airline                             & \checkmark                 &            0.01\%            & \checkmark             \\ \cline{2-6} 
                                                                                       & http://tinyurl.com/perthliv                  & Airport                             & \checkmark                 &             0.01\%           & \checkmark              \\ \hline
\multirow{5}{*}{\begin{tabular}[c]{@{}l@{}}Marine \\ traffic \\ tracking\end{tabular}} & http://www.marinetraffic.com                 & \multirow{3}{*}{Worldwide}          & \checkmark                 &                       100\% & -              \\ \cline{2-2} \cline{4-6} 
                                                                                       & http://www.shipspotting.com/ais/             &                                     & \checkmark                 &              0.01\%        & \checkmark              \\ \cline{2-2} \cline{4-6} 
                                                                                       & http://ship.gr/map/index.htm                 &                                     &  $\times$                 &              100\%         & $\times$              \\ \cline{2-6} 
                                                                                       & http://www.cruisemapper.com                  & \multirow{2}{*}{Cruise}             & \checkmark                 &          0.01\%            & $\times$    \\ \cline{2-2} \cline{4-6} 
                                                                                       & http://www.cruisin.me/cruise-ship-tracker/   &                                     & $\times$                &              100\%          & $\times$              \\ \hline
\end{tabular}
\end{center}
\end{small}
\vspace{-4mm}
\end{table*}

For the captured flight data, the objects information from different sources 
is quite different. There are two main reasons associated with this issue. The first reason is the delay in updating the information. The second reason is the loss of some values for some flights. For example, the flight registration is provided by a data source while the same attribute for the same flight is not set in other websites. For the marine traffic tracking websites, we observe that the majority of niche websites are using the same techniques and data as the source website. No delay is observed while the ratio of overlapped data is higher than flight trackers.

\subsection{IoT vs. User Interests}
As mentioned, the analysis of the distribution of things and queries can lead to finding more efficient strategies for storing and retrieving IoT data. 

Our observation shows that in many cases, there is a huge difference between the distribution of the queries and the distribution of the things data. Figures \ref{fig:asiaqmap}, \ref{fig:ausqmap} and \ref{fig:euqmap} show the local distribution of the queries from Thingful in Asia, Australia and the Europe, respectively. 
We do not include other continents such as the Americas and the Africa as their results do not add new information on top of the selected regions. 
As the figures show, in each region most of the queries are focused on specific regions such as India, East Coast of Australia and London. 

We also investigate the distribution of the things and its changes over the time in each region separately. We randomly pick a 12-hour time frame and conduct the analysis over three snapshots all over the world. 
Due to the space limit, we select three snapshots and investigate the distribution of the things in each region. The first snapshot is during evening, the second is during early morning  and the third is around noon. The snapshots are all based on the Australian Central Standard Time (ACST). 
In Asia, during the afternoon most of the things are concentrated on East Asia (Figure \ref{fig:asia1}) while later in the morning the concentration of the things transfers to the South West Asia (Figure \ref{fig:asia2}). In the next snapshot, the concentration moves towards South East Asia again (Figure \ref{fig:asia3}). A large part of this change is due to the existing large ratio of flight data comparing to the other types of things data in Asia. However, in Asia, no record demonstrates a good match between the distribution of things and the distribution of the queries, while most queries are concentrated on India (see Figure \ref{fig:asiaqmap}). 

In Australia, things are mostly concentrated on the east coast of Australia during evening time (Figure \ref{fig:aus1}) which is a good match for the distribution of the queries (Figure \ref{fig:ausqmap}). 
Later as Figures \ref{fig:aus2} and \ref{fig:aus3} show, many things are present in other places as well as around the capital cities of Sydney and Melbourne. 
Thus, a huge gap exists between the distribution of things and queries in this region. However, unlike 
Asia, 
the distribution of the things 
partially 
matches with the distribution of the queries over the two cities. 


The situation is quite different in Europe. As Figures \ref{fig:europe1}, \ref{fig:europe2} and \ref{fig:europe3} show, a large number of things are constantly concentrated over London area and partially over 
Germany which is a very good match with the distribution of the queries in this part of the world (Figure \ref{fig:euqmap}).


\section{Discussions}
\label{sec:discussion}
In this section, we provide 
further discussions on the challenges and opportunities for IoT research and development.

\subsection{Challenges in IoT Data Discovery}
IoT data discovery is 
important 
towards establishing the next step in the life of the Internet. Since the early days of IoT, several technologies have been specifically proposed to share IoT data. There have been 
several successful stories for cloud based IoT platforms such as Xively and Paraimpu.
They are often designed to provide global support for almost any type of sensors or actuators.

However, the community of users 
does not restrict themselves to what these IoT platforms provide. An increasing number of techniques are being used to publish sensory data on the Web. The number of sensors, various types of applications and the increasing demand for real time data have driven 
to reinvent various techniques which 
are previously 
used for other purposes (e.g., Web Mapping). In this case, a large number of niche websites 
have been developed to publish the data that 
are generated by specific sensors or for specific applications. In fact, the volume of the publicly available information provided by these websites is much more than the general purpose IoT cloud platforms. However, identifying these 
websites 
is similar to finding a needle in haystack as there is no comprehensive list of such websites, many of which have been created recently after the success of similar applications such as the flight trackers.

Another challenge is the structure of the data that is provided on the Web. For a large portion of the websites, the data should be collected from the deep Web. For example, to obtain the results from a flight tracker website, several parameters need to be set and passed. Otherwise, a small subset of the data or an empty set will be provided by the server. In some cases, authentication may also be required as a part of the process 
when accessing the data. 

Unlike other types of the information on the Web, IoT data mostly are presented in a structured or semi-structured format. The structure of the data widely varies from one website to another. In addition, in many cases, the parameter names are not self descriptive and these parameters need to be demystified manually.

\subsection{Information Retrieval in IoT}
Currently, IoT search is quickly evolving to address the needs of users and leverage the benefits of deploying IoT. This includes preserving accuracy, speed, consistency and ease of use for IoT search engines in future. Thus, new search forms such as searching the retrieving knowledge from things data, intent-based search are emerging which in turn require proper data source selection, tackling query ambiguity \cite{shemshadi2015ecs}. 

We consider a simple example to further explain this point. For example, a user may search for air quality in a specific area rather than a specific air sensor. To answer her query, firstly, the documents containing air quality sensors in that  area should be retrieved. Secondly, sensors which also provide contextual information about the air quality should be retrieved as well. Thirdly, knowledge about air quality can be extracted from the selected documents. We should note that due to the high uncertainty in IoT data, some estimation or prediction (in case of data unavailability) techniques may be used to fill the empty pieces of the puzzle. Finally, result diversification would be very important to address issues such as query ambiguity. 

Due to the highly dynamic nature of IoT, documents containing IoT information can change drastically from the time 
when they are crawled to the time 
when their data are presented. Thus, effective and efficient indexing techniques would be required to retrieve information from IoT data. 

Furthermore, IoT will leverage \textit{Temporal Information Retrieval} more \cite{kanhabua2015temporal}. We observe that the content of a large number of documents as well as the data sources, may variate between 1\% to 100\% in a very short period of time. With the high rate of changes in IoT data sources, document selection as well as the user query results will require novel temporal techniques to tackle the issues.

\subsection{Other Challenges}

Our dataset can be used for a variety of purposes in the IoT research and development, including correlation discovery between things \cite{yao2013correlation}, IoT data storage \cite{ma2012efficient,jiang2014iot}, context aware computing for IoT \cite{perera2013context} by merging sensor readings from different sources such as environmental and transportation sensors, point of interest recommendation \cite{yao2015point} and other active IoT research areas which may need real-world data.

\subsubsection{Data Integration}
Continuous retrieval of IoT data is very challenging. Some of the sources demand authentication before providing the 
access. 
In many cases, data for the same object (e.g., an aircraft) is being broadcasted by different data sources. Furthermore, in some cases, each available source may only provide 
partial information for an object.
The similar issue affects merging data for the same resource. For example, the results of parsing objects data for a single resource have different length and parameters which need to be integrated at the end. Lastly, many data sources limit their response length due to load balancing concerns. 
We do not fully resolve all challenges in the integration but rather, we use an efficient approach to integrate the data from different sources for the purpose of our research. However, the integration of IoT data is more challenging than what is believed and more research in this area is required in the future. 

\subsubsection{Scalability}
Collecting, processing and storing IoT data can be a time consuming procedure, particularly if the size of the dataset is large. As the number of sources and objects increases, which might count in billions, using one instance of the crawler would be very inefficient. In particular, dramatic difference between the update rate of different data sources which also partially depends on their size, can be challenging. 
Furthermore, technical failures of one resource may affect collecting data from other data sources in the same chain. Thus, we use a distribution strategy to coordinate different instances of the crawler running on different machines.

\subsubsection{Archiving IoT Data}
IoT fully interprets the Big Data. The volume, velocity and the variety of the data generated by things are enormous. The amount of the data that is already being published from 20 IoT data sources on the Web, which we estimate to be more than 100 TB a day, is already comparable to the amount of data that is being generated by users on social networks. With the rapid growth of the 
IoT, in the near future, new techniques will be required to effectively and efficiently process and store IoT data. 

Currently, to the best of our knowledge, there is no popular website for archiving the publicly available IoT data.
In this regard, the traditional approaches need to be revised for the new era of the IoT. The result can be valuable to many core applications such as IoT search while we compromise on some issues to make the impossibles possible. For instance, we discover that based on the changes in the geographical distribution of objects, a crawling strategy can be issued to capture the most updated data in the least amount of time. Through creating spatial and per resource indexes, the process can also become more optimized.


\section{Related Work}
\label{sec:relatedwork}

Over the past few years, the IoT has received increasing attention from researchers and practitioners. 
In the earlier days, Atzori et al.\cite{atzori2010internet} offers an initial survey on the IoT research. 
Accordingly, there exist manifold definitions of the IoT paradigm within the research community. Each definition may view this paradigm from a specific angle including things oriented, Internet oriented and semantic oriented definitions. More specifically, based on all these definitions, a wide variety of networking and sharing technologies have been used to enable the future IoT including but not limited to the Web of Things (WoT) \cite{raggett2015web}, RFID, Near-Field Communication (NFC), middleware and the Wireless Identification and Sensing Platform (WISP)\footnote{http://wisp.wikispaces.com/}. 

Some researchers have claimed that the IoT is implemented with the technologies specifically designed for the purpose of being deployed in IoT \cite{want2015enabling}. Thus, it is argued that the IoT already exists but only a small number of experiments and as a result, yet many researchers consider as inaccessible \cite{want2015enabling}. Restricting IoT with this point of view is contrary to the spirit of open systems at the heart of the original Internet standards. Moreover, within the technologies which have been applied to facilitate IoT, open Web technologies including HTML, Ajax, HTTPS, OpenID and structured data apply equally well to IoT. However, yet there is no advanced mechanism to be able to effectively search and retrieve things from the Web. 

The very diverse range of the objects, approaches and technologies used to implement IoT have contributed in broadening the definition of this paradigm. For instance, the IoT can be realized through deploying an RFID ecosystem consisting of objects tagged with numerous RFID labels \cite{welbourne2009building}. Another option is to build the IoT using smart objects \cite{kortuem2010smart} which in turn can be divided into activity-aware, policy-aware and process-aware smart objects.


Currently IoT search is a trending research direction \cite{whitmore2014internet} with stress over some major goals such as real-time search~\cite{ostermaier2010real}, context-awareness  \cite{perera2013context} and relationship support \cite{nitti2014friendship}. 

Researchers have complained about the lack of real-world IoT data in the past \cite{yao2013correlation}. Although some of the previous works have claimed testing their proposed solutions for large scale IoT data, such as meta-heuristic \cite{ebrahimi2015new} or context aware sensor search \cite{perera2013context}, all 
these previous works mainly deal with small or simulated datasets. To the best of our knowledge, no work has ever collected or analyzed large scale things data. 
In addition, none of the existing works use real IoT search query dataset. 
Moreover, we could not find any other work that has deployed or analyzed a large real-world IoT query dataset for mining user interests in the IoT domain. By combining the user interests and the 
IoT data, we can analyze the gap of what people look for and what currently the IoT presents on the Web. 
Our work is the very first that investigates IoT in large scale and the dataset released from our study is the first real-life, large IoT dataset that is publicly available for the research community. 

\section{Conclusion}
\label{sec:conclusion}
The unique characteristics of IoT require that the future search engines provide support for a variety of new requirements.
Firstly, IoT is no longer solely attached to the public cloud based solutions, and IoT indeed has been spreading by all means throughout the World Wide Web. 
Secondly, due to the dynamics of IoT, future search engines need to provide real-time results.
In this work, we conduct an in-depth analytical investigation on IoT data that exists on the Web. 
Based on our real-life IoT data, we investigate the current status of the IoT and identify open research and development issues.
Our findings show that conventional geospatial data presentation techniques such as Web Mapping play an important role in distributing IoT data. Thus, to find publicly available IoT data, one should look for pages with live and interactive maps. 
%
For the future work, we plan to investigate more sophisticated crawling strategies to achieve more comprehensive collection of IoT data 
for real-time search and analysis of IoT data. 
We also plan to use the collected IoT data for developing novel 
applications such as real-time prediction of flight delays.

\bibliographystyle{abbrv}
\bibliography{cikm16}

\begin{thebibliography}{10}

\bibitem{atzori2010internet}
L.~Atzori, A.~Iera, and G.~Morabito.
\newblock The internet of things: A survey.
\newblock {\em Computer networks}, 54(15):2787--2805, 2010.

\bibitem{blackstock2012iot}
M.~Blackstock and R.~Lea.
\newblock Iot mashups with the wotkit.
\newblock In {\em Proc. of the 3rd Intl. Conf. on the Internet of Things (IOT
  2012)}, pages 159--166. IEEE, 2012.

\bibitem{castro2012analysis}
M.~Castro, A.~J. Jara, and A.~F. Skarmeta.
\newblock An analysis of m2m platforms: challenges and opportunities for the
  internet of things.
\newblock In {\em Proc. of the 6th Intl. Conf. on Innovative Mobile and
  Internet Services in Ubiquitous Computing (IMIS 2012)}, pages 757--762. IEEE,
  2012.

\bibitem{da2014internet}
L.~Da~Xu, W.~He, and S.~Li.
\newblock Internet of things in industries: A survey.
\newblock {\em IEEE Transactions on Industrial Informatics}, 10(4):2233--2243,
  2014.

\bibitem{ebrahimi2015new}
M.~Ebrahimi, E.~Shafieibavani, R.~K. Wong, and C.-H. Chi.
\newblock A new meta-heuristic approach for efficient search in the internet of
  things.
\newblock In {\em Proceedinsg of the 12th IEEE Intl. Conf. on Services
  Computing (SCC 2015)}, pages 264--270. IEEE, 2015.

\bibitem{gubbi2013internet}
J.~Gubbi, R.~Buyya, S.~Marusic, and M.~Palaniswami.
\newblock Internet of things (iot): A vision, architectural elements, and
  future directions.
\newblock {\em Future Generation Computer Systems}, 29(7):1645--1660, 2013.

\bibitem{haklay2008web}
M.~Haklay, A.~Singleton, and C.~Parker.
\newblock Web mapping 2.0: The neogeography of the geoweb.
\newblock {\em Geography Compass}, 2(6):2011--2039, 2008.

\bibitem{jiang2014iot}
L.~Jiang, L.~Da~Xu, H.~Cai, Z.~Jiang, F.~Bu, and B.~Xu.
\newblock An iot-oriented data storage framework in cloud computing platform.
\newblock {\em IEEE Transactions on Industrial Informatics}, 10(2):1443--1451,
  2014.

\bibitem{kanhabua2015temporal}
N.~Kanhabua, R.~Blanco, K.~N{\o}rv{\aa}g, et~al.
\newblock Temporal information retrieval.
\newblock {\em Foundations and Trends{\textregistered} in Information
  Retrieval}, 9(2):91--208, 2015.

\bibitem{kim2014m2m}
J.~Kim, J.~Lee, J.~Kim, and J.~Yun.
\newblock M2m service platforms: survey, issues, and enabling technologies.
\newblock {\em Communications Surveys \& Tutorials, IEEE}, 16(1):61--76, 2014.

\bibitem{kortuem2010smart}
G.~Kortuem, F.~Kawsar, D.~Fitton, and V.~Sundramoorthy.
\newblock Smart objects as building blocks for the internet of things.
\newblock {\em IEEE Internet Computing}, 14(1):44--51, 2010.

\bibitem{ma2012efficient}
Y.~Ma, J.~Rao, W.~Hu, X.~Meng, X.~Han, Y.~Zhang, Y.~Chai, and C.~Liu.
\newblock An efficient index for massive iot data in cloud environment.
\newblock In {\em Proc. of the 21st ACM Intl. Conf. on Information and
  Knowledge Management (CIKM 2012)}, pages 2129--2133, 2012.

\bibitem{nitti2014friendship}
M.~Nitti, L.~Atzori, and I.~Cvijikj.
\newblock Friendship selection in the social internet of things: Challenges and
  possible strategies.
\newblock {\em Internet of Things Journal, IEEE}, 2(3):240--247, June 2015.

\bibitem{ostermaier2010real}
B.~Ostermaier, K.~Römer, F.~Mattern, M.~Fahrmair, and W.~Kellerer.
\newblock A real-time search engine for the web of things.
\newblock In {\em Proc. of the 2nd Internet of Things Conf. (IOT 2010)}, pages
  1--8. IEEE, 2010.

\bibitem{pang2013ecosystem}
Z.~Pang, Q.~Chen, J.~Tian, L.~Zheng, and E.~Dubrova.
\newblock Ecosystem analysis in the design of open platform-based in-home
  healthcare terminals towards the internet-of-things.
\newblock In {\em Proc. of the 15th Intl. Conf. on Advanced Communication
  Technology (ICACT 2013)}, pages 529--534, 2013.

\bibitem{perera2013context}
C.~Perera, A.~Zaslavsky, P.~Christen, M.~Compton, and D.~Georgakopoulos.
\newblock Context-aware sensor search, selection and ranking model for internet
  of things middleware.
\newblock In {\em Proc. of the 14th IEEE Intl. Conf. on Mobile Data Management
  (MDM 2013)}, volume~1, pages 314--322. IEEE, 2013.

\bibitem{pintus2012paraimpu}
A.~Pintus, D.~Carboni, and A.~Piras.
\newblock Paraimpu: a platform for a social web of things.
\newblock In {\em Proc. of the 21st Intl. Companion Conf. on World Wide Web
  (WWW 2012)}, pages 401--404, 2012.

\bibitem{Qin-JNCA16}
Y.~Qin, Q.~Z. Sheng, N.~J.~G. Falkner, S.~Dustdar, H.~Wang, and A.~V.
  Vasilakos.
\newblock {When Things Matter: A Survey on Data-Centric Internet of Things}.
\newblock {\em Journal Network and Computer Applications}, 64:137--153, 2016.

\bibitem{qinSFDWV2014}
Y.~Qin, Q.~Z. Sheng, and W.~E. Zhang.
\newblock {SIEF: Efficiently Answering Distance Queries for Failure Prone
  Graphs}.
\newblock In {\em Proc. of the 18th Intl. Conf. on Extending Database
  Technology (EDBT)}, pages 145--156, 2015.

\bibitem{raggett2015web}
D.~Raggett.
\newblock The web of things: Challenges and opportunities.
\newblock {\em Computer}, 48(5):26--32, 2015.

\bibitem{shemshadi2015ecs}
A.~Shemshadi, L.~Yao, Y.~Qin, Q.~Z. Sheng, and Y.~Zhang.
\newblock Ecs: A framework for diversified and relevant search in the internet
  of things.
\newblock In {\em Web Information Systems Engineering--WISE 2015}, pages
  448--462. 2015.

\bibitem{emdist2015}
S.~Urbanek and Y.~Rubner.
\newblock {\em emdist: Earth Mover's Distance}, 2012.
\newblock R package version 0.3-1.

\bibitem{want2015enabling}
R.~Want, B.~N. Schilit, and S.~Jenson.
\newblock Enabling the internet of things.
\newblock {\em Computer}, 48(1):28--35, 2015.

\bibitem{welbourne2009building}
E.~Welbourne, L.~Battle, G.~Cole, K.~Gould, K.~Rector, S.~Raymer,
  M.~Balazinska, and G.~Borriello.
\newblock Building the internet of things using rfid: the rfid ecosystem
  experience.
\newblock {\em IEEE Internet Computing}, 13(3):48--55, 2009.

\bibitem{whitmore2014internet}
A.~Whitmore, A.~Agarwal, and L.~Da~Xu.
\newblock The internet of things--a survey of topics and trends.
\newblock {\em Information Systems Frontiers}, 17(2):261--274, 2014.

\bibitem{wu2011m2m}
G.~Wu, S.~Talwar, K.~Johnsson, N.~Himayat, and K.~D. Johnson.
\newblock M2m: From mobile to embedded internet.
\newblock {\em IEEE Communications Magazine}, 49(4):36--43, 2011.

\bibitem{yao2013correlation}
L.~Yao and Q.~Z. Sheng.
\newblock Correlation discovery in web of things.
\newblock In {\em Proc. of the 22nd Intl. Comp. Conf. on World Wide Web (WWW
  2013)}, pages 215--216, 2013.

\bibitem{yao2015point}
L.~Yao, Q.~Z. Sheng, Y.~Qin, X.~Wang, A.~Shemshadi, and Q.~He.
\newblock {Context-aware Point-of-Interest Recommendation Using Tensor
  Factorization with Social Regularization}.
\newblock In {\em Proc. of the 38th Intl. ACM SIGIR Conf. on Research and
  Development in Inf. Retrieval (SIGIR)}, pages 1007--1010, 2015.

\end{thebibliography}

\end{document}